\documentclass[10pt,aps,twocolumn,prd,notitlepage,amssymb,amsmath,floatfix,nofootinbib,superscriptaddress]{revtex4-1}

\usepackage{epsfig}
\usepackage{bm}
\usepackage{amssymb}
\usepackage{amsmath}
\usepackage{color}
\usepackage{subfigure}
\usepackage[colorlinks,linkcolor=blue,anchorcolor=black,citecolor=blue]{hyperref}

\mathchardef\mhyphen="2D

\begin{document}

\title{The Spin Alignment of Vector Mesons in High Energy $pp$ Collisions}

\author{Kai-bao Chen}
\affiliation{School of Science, Shandong Jianzhu University, Jinan, Shandong 250101, China}

\author{Zuo-tang Liang}
\affiliation{Institute of Frontier and Interdisciplinary Science, 
Key Laboratory of Particle Physics and Particle Irradiation (MOE), Shandong University, Qingdao, Shandong 266237, China}

\author{Yu-kun Song}
\affiliation{School of Physics and Technology, University of Jinan, Jinan, Shandong 250022, China}

\author{Shu-yi Wei}
\affiliation{European Centre for Theoretical Studies in Nuclear Physics and Related Areas (ECT*)\\
and Fondazione Bruno Kessler, Strada delle Tabarelle 286, I-38123 Villazzano (TN), Italy}

\begin{abstract}
The spin alignment of vector mesons produced in high energy reactions is determined by 
the spin-dependent fragmentation function $D_{1LL} (z,\mu_f)$ that is shown to be independent of 
the polarization of the fragmenting quark.  
In this paper, we extract the spin-dependent fragmentation function $D_{1LL} (z,\mu_f)$
from data on the spin alignment of $K^{*0}$ in $e^+e^-$ annihilation at LEP in two different scenarios
and apply them to make predictions in $pp$ collisions. 
We make detailed analysis of contributions from different sub-processes and show that 
the spin alignment should be quite significant also in high energy $pp$ collisions. 
\end{abstract}

\maketitle

\section{Introduction}

The spin dependence of fragmentation functions (FFs) is one of the important aspects in high energy spin physics and 
plays an important role in studying the properties of Quantum Chromodynamics (QCD) in general 
and the hadronization mechanism in particular.
So far as the polarization of produced hadrons is concerned, 
two classes of polarizations have been often studied, i.e., the vector and the tensor polarizations.
The former can be studied by measuring the polarization of hyperons via their spin self analyzing weak decays, 
and the latter is studied via strong decays of vector mesons into two pseudo-scalar mesons. 
The tensor polarization is usually decomposed into five components. 
Among them, the $S_{LL}$-component is directly related to the probability 
for the third component of spin to take zero that is called the spin alignment. 
The spin alignment of vector mesons has been measured in $e^+e^-$-annihilations 
and other high energy reactions~\cite{Abreu:1997wd,Ackerstaff:1997kj,Ackerstaff:1997kd,Abbiendi:1999bz,Chukanov:2006xa,Abelev:2008ag,Zhou:2019lun,Acharya:2019vpe}. 

Comparing with that of parton distribution functions (PDFs), 
we know even less about the spin dependence of FFs.
Among different aspects, hyperon polarizations are best studied
both experimentally~\cite{Lesnik:1975my,Bunce:1976yb,Bensinger:1983vc,Gourlay:1986mf,Adamovich:1994gy,
Althoff:1984iz,Buskulic:1996vb,Ackerstaff:1997nh,
Airapetian:1999sh,Airapetian:2006ee,Astier:2000ax,Astier:2001ve,Alekseev:2009ab,Abelev:2009xg,Adam:2018kzl,Adam:2018wce,Guan:2018ckx,BESIII}  
and phenomenologically~\cite{Gustafson:1992iq,
Liang:1997rt,Boros:1998kc,Liu:2000fi,Liu:2001yt,Zuotang:2002ub,Xu:2002hz,Dong:2005ea,Xu:2005ru,Chen:2007tm,Zhou:2008fb,Zhou:2009mx,
Ma:1998pd,Ma:1999gj,Ma:1999wp,Ma:1999hi,Ma:2000uu,Ma:2000cg,Chi:2013hka,Liu:2019xcf,Ellis:2002zv}.
Parameterizations of the corresponding spin dependent FFs have been proposed~\cite{deFlorian:1997zj}. 

For the tensor polarization of vector mesons, the study has in fact two advantages: 
(1) there is little contamination from decay processes; 
(2) no decay parameter is involved in the two-body strong decay of the vector meson 
so that there is no uncertainty caused by the decay parameter~\cite{BESIII} 
and the measurement efficiency is high.  
Measurements have been carried out on the spin alignment and also  
the off-diagonal components in high energy reactions~\cite{Abreu:1997wd,Ackerstaff:1997kj,Ackerstaff:1997kd,Abbiendi:1999bz,Chukanov:2006xa,Abelev:2008ag,Zhou:2019lun,Acharya:2019vpe}.  
We have in particular data on the spin alignment with relatively high accuracy from 
experiments at LEP~\cite{Abreu:1997wd,Ackerstaff:1997kj,Abbiendi:1999bz,Ackerstaff:1997kd}.
The data show an evident spin alignment of vector mesons produced in $e^+e^-$ annihilations 
and triggered many phenomenological studies~\cite{Anselmino:1984af,Anselmino:1997ui,Anselmino:1998jv,Anselmino:1999cg,Xu:2001hz,Xu:2002vz,Xu:2003fq,Xu:2003rs}. 
Since the collision energy is at the $Z^0$ pole, the fragmenting quark and anti-quark are highly polarized. 
Therefore, it was quite natural to attribute the spin alignment to the polarization of the parent quark and/or anti-quark. 
Most of the phenomenological efforts have been accomplished following such a perception~\cite{Xu:2001hz,Xu:2002vz,Xu:2003fq,Xu:2003rs}. 

Recently, progresses in the theoretical study
have been achieved in particular in the formal QCD description of the spin dependence of 
FFs~\cite{Boer:1997mf,Boer:1997qn,Boer:2008fr,Pitonyak:2013dsu,Wei:2013csa,Wei:2014pma,Chen:2015ora,Chen:2016moq,
Chen:2016iey,Wei:2016far,Chen:2015tca}. 
In the QCD field theory, FFs are defined via Lorentz decompositions of the quark-quark correlator. 
A systematic study of such a decomposition has been accomplished~\cite{Chen:2015ora,Chen:2016moq} and 
the results show in particular that the spin alignment is determined solely by 
the $S_{LL}$-dependent FF $D_{1LL}$ and $D_{1LL}$ is independent of the spin of the fragmenting quark. 
Correspondingly the first attempt to extract $D_{1LL}(z)$ from the LEP data~\cite{Abreu:1997wd,Ackerstaff:1997kj,Abbiendi:1999bz} has been performed in~\cite{Chen:2016iey}.

Although it might be counter-intuitive, this conclusion is actually expected by the parity invariance. 
This can be seen clearly in the helicity base. 
As a component of the polarization tensor, $S_{LL}$ is a scalar that is invariant under the space inversion. 
Hence, one cannot establish a relation between $S_{LL}$ and the helicity of the quark in a parity conserved manner. 
This is quite different from the case for the longitudinal polarization of $\Lambda$, 
where $\lambda_q \lambda_{\Lambda}$ is a parity-invariant structure that should be included 
in the decomposition of fragmentation function, 
where $\lambda_q$ and $\lambda_\Lambda$ are helicities of the quark and $\Lambda$ respectively. 

Though the prediction is very solid, it is however quite difficult to understand 
why the fragmentation of an unpolarized quark leads to vector meson 
with a larger probability at the helicity zero state. 
Experimental check of the quark polarization independence of the vector meson spin alignment 
should be a very basic test of the fragmentation picture  
and deep studies in this direction should lead to new insights on the hadronization mechanism. 
In this connection, it might be also interesting to mention that spin effects have also attracted 
much attention recently in heavy ion collisions. 
Here, a very special state of hadronic matter -- the quark gluon plasma (QGP) is formed 
and the hadronization mechanism is different. 
Both hyperon polarization and vector meson spin alignment have been studied at RHIC as well as at the LHC in this connection.
The studies have been inspired by the theoretical predictions~\cite{Liang:2004ph,Liang:2004xn} 
and the experimental confirmation~\cite{STAR:2017ckg} on the global polarization of QGP with respect to the reaction plane. 
The vector meson spin alignment was predicted~\cite{Liang:2004xn} to be strongly dependent on the global polarization of quarks and anti-quarks because they are produced via the quark coalescence rather than the quark fragmentation mechanism.

Currently, both RHIC and the LHC provide good opportunities in experiments to study vector meson spin alignment in $pp$ collisions. 
In particular at RHIC the quark polarization independence can easily be tested since RHIC is also a polarized $pp$ collider. 
It is thus timely and important to make predictions for such measurements. 
 
In this paper, we study the spin alignment of vector mesons in $pp\to VX$. 
We extract the $S_{LL}$-dependent FF $D_{1LL}$ from the LEP data and make predictions for $pp$ collisions. 
In Sec.~II, we present the basic formulae needed for such numerical calculations.
In Sec.~III, we present parameterizations of $D_{1LL}$ and numerical results in Sec.~IV. 
A short summary is given in Sec.~V.

\section{The formalism}

In this section, we present the differential cross section of vector meson production in $pp$ collisions 
needed to calculate the spin alignment. 
We do the calculations up to the order where the first order of pQCD evolution of FFs is included, 
and present the formulae needed for such calculations. 

\subsection{The differential cross section}

We consider $pp\to VX$ in the high $p_T$ region where collinear factorization is applicable 
and study the spin alignment of produced vector meson $V$.
Since the spin alignment is independent of the polarization of the fragmenting quark, 
the calculations are similar in the polarized or unpolarized collisions. 
We simply take unpolarized $pp$ as an example. 

To calculate the spin alignment of $V$, we need to consider the spin dependent differential cross section.
We recall that the polarization of spin-1 particles is described by a $3\times 3$ spin density matrix $\rho$.
In the rest frame of the particle, $\rho$ is usually decomposed as~\cite{Bacchetta:2000jk,Chen:2015ora,Chen:2016moq},
\begin{align}
\rho = \frac{1}{3} (\mathbf{1} + \frac{3}{2}S^i \Sigma^i + 3 T^{ij} \Sigma^{ij}), \label{eq:spin1rho}
\end{align}
where $\Sigma^i$ is the spin operator of a spin-$1$ particle, 
and $\Sigma^{ij}= \frac{1}{2} (\Sigma^i\Sigma^j + \Sigma^j \Sigma^i) - \frac{2}{3} \mathbf{1} \delta^{ij}$.
$T^{ij}={\rm Tr}(\rho \Sigma^{ij})$ is the polarization tensor and is parameterized as,
\begin{align}
\mathbf{T}= \frac{1}{2}
\left(
\begin{array}{ccc}
-\frac{2}{3}S_{LL} + S_{TT}^{xx} & S_{TT}^{xy} & S_{LT}^x  \\
S_{TT}^{xy}  & -\frac{2}{3} S_{LL} - S_{TT}^{xx} & S_{LT}^{y} \\
S_{LT}^x & S_{LT}^{y} & \frac{4}{3} S_{LL}
\end{array}
\right).
\label{eq:spintensor}
\end{align}
Here, the polarization vector $S$ is similar to that for spin-1/2 hadrons. 
The polarization tensor $T$ is further decomposed into a Lorentz scalar $S_{LL}$, 
a Lorentz vector $S_{LT}^\mu = (0, S_{LT}^x, S_{LT}^y,0)$, 
and a Lorentz tensor $S_{TT}^{\mu\nu}$ that has two nonzero independent components 
$S_{TT}^{xx} = -S_{TT}^{yy}$ and $S_{TT}^{xy} = S_{TT}^{yx}$. 
It has in total five independent components.
The spin alignment $\rho_{00}$ is directly related to $S_{LL}$ by $\rho_{00}=(1-2S_{LL})/3$, 
where $\rho_{00}$ takes the physical meaning of the probability for the third component $m$ of spin of $V$ to take zero 
while $S_{LL}=(\rho_{++}+\rho_{--})/2-\rho_{00}$ is the difference of $m$ to take $\pm 1$ and $0$. 
In the helicity basis, $m$ is just the helicity $\lambda_V$ of the vector meson $V$.

To calculate the spin alignment $\rho_{00}$ of the produced vector meson $V$, 
we need to consider the $S_{LL}$-dependent part of the cross section and sum over all other components of polarization. 
Since $S_{LL}$ is a Lorentz scalar, the $S_{LL}$-dependent part takes the same form as that of the unpolarized part.
In this way, we obtain the differential cross section in the collinear factorization form as~\cite{Owens:1986mp}, 
\begin{align}
&\frac{d\sigma_{pp\to VX}}{dyd^2p_T} = \sum_{abcd} \int d y_2 \int \frac{dz}{z^2} x_1 f_a (x_1,\mu_f) x_2 f_b(x_2,\mu_f) \nonumber\\
&~~~\times \frac{1}{\pi} \frac{d\hat\sigma_{ab\to cd}}{d\hat t} [D_{1c}^V (z,\mu_f) + S_{LL} D_{1LLc}^V (z, \mu_f)], 
\label{eq:x-pp-vX}
\end{align}
where $f_{a,b} (x_{i},\mu_f)$ is the parton distribution function~\cite{Dulat:2015mca} with $x_i$ the longitudinal momentum fraction 
and $\mu_f$ the factorization scale, $D_{1c}^V(z,\mu_f)$ and $D_{1LLc}^V(z,\mu_f)$ are 
the spin averaged and $S_{LL}$-dependent FFs of $c\to VX$ respectively; 
$y$ and $p_T$ denote the rapidity and transverse momentum of $V$  
and they are related to $x_1, x_2$ and $z$ by $x_1 = {p_T} (e^{y} + e^{y_2}) / {z}\sqrt{s}$,
$x_2 = {p_T} (e^{-y} + e^{-y_2}) / {z}\sqrt{s}$; $y_2$ is the rapidity of parton $d$ after the scattering;  
$d\hat\sigma_{ab\to cd}/d\hat t$ is the cross section of the partonic process $ab\to cd$ at the leading order.
The partonic process includes all the elementary processes at the parton level
such as $q_1q_2\to q_1q_2$, $q_1\bar q_2\to q_1\bar q_2$, $q_1q_1\to q_1q_1$, 
$q_1g\to q_1g$, $gg\to gg$, $q_1\bar q_1\to q_1\bar q_1$,  $q_1\bar q_1\to q_2\bar q_2$, 
$q\bar q\to gg$, and $gg\to q\bar q$. 
We consider the unpolarized reaction and the cross sections for these elementary processes are available in literature~\cite{Owens:1986mp}.
Here, we note in particular that in Eq.~(\ref{eq:x-pp-vX}), FFs are defined for a given polarization state following 
the same convention as that in~\cite{Wei:2013csa}  
where $D_{1c}^V(z,\mu_f)$ is the spin-averaged FF and is related to the spin-summed FF $D_c^V (z,\mu_f)$ by $D_c^V (z,\mu_f) = 3D_{1c}^V(z,\mu_f)$.

Besides presenting the differential cross section in terms of $y$ and $p_{T}$, 
we can also make predictions in terms of other variables such as $(x_F,p_T)$ 
where $x_F \equiv 2p_{z}/\sqrt{s}= {2m_T\sinh y}/{\sqrt{s}}$, $m_T=\sqrt{m^2+p^2_T}$, and
\begin{align}
& dy d^2p_T 
= dx_Fd^2p_T/{\sqrt{x_F^2+4m_T^2/s}}.
\end{align}

\subsection{The spin alignment}

The spin-alignment of $V$ is then given by,
\begin{align}
\rho_{00}^V = {d\sigma^{\lambda_V = 0}}\Big/\sum_{\lambda_V=\pm1,0}d\sigma^{\lambda_V}. \label{eq:rhodef}
\end{align}

For the helicity $\lambda_V = \pm 1$ state, $S_{LL} = {1}/{2}$, while for $\lambda_V = 0$ state, $S_{LL} = -1$. 
Hence, we obtain,
\begin{align}
  \rho_{00}^V(y,p_T)&=\frac{1}{3}-\frac{d\sigma_{pp\to VX}^{\rm S_{LL}}}{dyd^2p_T}\Big/
                      \frac{d\sigma_{pp\to VX}^{\rm spin \mhyphen summed}}{dyd^2p_T},\label{eq:rho00}
\end{align}
where the spin-summed cross section 
is given by,
\begin{align}
&\frac{d\sigma_{pp\to VX}^{\rm spin \mhyphen summed}}{dyd^2p_T} 
= 3 \sum_{abcd} \int d y_2 \int \frac{dz}{z^2} x_1 f_a (x_1,\mu_f)  \nonumber\\
&~~~~~~~\times x_2 f_b(x_2,\mu_f)\frac{1}{\pi}\frac{d\hat\sigma_{ab\to cd}}{d\hat t} D_{1c} (z,\mu_f), \label{eq:cs-spinsummed}
\end{align}
while the $S_{LL}$-dependent part is,
\begin{align}
\frac{d\sigma_{pp\to VX}^{\rm S_{LL}}}{dyd^2p_T} 
&=  \sum_{abcd} \int d y_2 \int \frac{dz}{z^2} x_1 f_a (x_1,\mu_f) \nonumber\\
&\times x_2 f_b(x_2,\mu_f) \frac{1}{\pi}\frac{d\hat\sigma_{ab\to cd}}{d\hat t}  D_{1LLc}^V (z, \mu_f).\label{eq:csSLL}
\end{align}

From the definition of $S_{LL}$ in particular its relation to $\rho_{00}$ we see that its value range is $-1\le S_{LL}\le 1/2$
so that $-2\le D_{1LL} (z,\mu_f)/D_{1} (z,\mu_f) \le 1$. 
In this way $0\le\rho_{00}\le 1$ is guaranteed.

\subsection{The QCD evolution of $D_{1LL}$}

The QCD evolution of collinear FFs is given by corresponding 
DGLAP equations~\cite{Dokshitzer:1977sg,Gribov:1972ri,Gribov:1972rt,Altarelli:1977zs} 
with time-like splitting functions~\cite{Owens:1978qz,Georgi:1977mg,Uematsu:1978yw}. 
The evolution equation of the $S_{LL}$-dependent FF $D_{1LL}$ is the same as that for unpolarized FF $D_1$, i.e., 
\begin{align}
&\frac{\partial}{\partial \ln Q^2} D_{1LLa}^{h} (z,Q^2)\nonumber\\
&~~~~~~=\frac{\alpha_s(Q^2)}{2\pi} \sum_b \int_z^1\frac{d\xi}{\xi}D_{1LLb}^{h}(\frac{z}{\xi},Q^2) P_{ba}(\xi),\label{eq:DGLAPD1LL}
\end{align}
where $a$ or $b$ denotes different types of partons including different flavors of quarks, anti-quarks and gluon.  
$P_{ba}(\xi)$ is just the leading order splitting function.

\section{The parameterization of the fragmentation function}
\label{sec:FF}

Even in the unpolarized case, we do not have an appropriate parameterization for the fragmentation function of vector mesons. 
Hence, we take the form of parameterizations based on symmetry properties, models and conjunctions and fix the free parameters 
using data available. 

\subsection{The unpolarized fragmentation function}
\label{sec:FFunp}
 
Currently, there is no parameterization of the fragmentation function of vector meson production available in the market even for the unpolarized case. 
However, we have parameterizations of the pseudo-scalar meson ($K^{\pm}$) production e.g. AKK08~\cite{Albino:2008fy} and DHESS~\cite{deFlorian:2017lwf}. 
Also a simple relation between the yields of $K^{*0}/\bar K^{*0}$ and $K^{\pm}$ has been observed~\cite{Shlyapnikov:2001jf} 
that leads to a linear dependence of $z$ for the ratio $D_{1u}^{K^{*+}}/D_{1u}^{K^+}$ approximately~\cite{Chen:2016iey}, i.e., 
\begin{align}
&D_{1u}^{K^{*+}} (z, \mu_0) = A (2z+1) D_{1u}^{K^+} (z, \mu_0), \label{eq:relKStarK+}
\end{align}
where $\mu_0 = 2$ GeV is the initial scale and $A\approx 0.3$ is the overall normalization factor. 
We extend this relation to FFs of all different kaons, i.e., 
\begin{align}
&D_{1a}^{K^{*}} (z, \mu_0) = A (2z+1) D_{1a}^{K} (z, \mu_0), \label{eq:relKStarK}
\end{align}
where $a$ stands for $u,d,s,\bar u,\bar d,\bar s$ and gluon $g$; 
$K^*$ stands for $K^{*\pm,0}$ and $\bar K^{*0}$ and $K$ for the corresponding pseudo-scalar mesons. 

For FFs of pseudo-scalar mesons, we apply isospin and charge conjugation symmetries and take, 
\begin{align}
&D_{1u}^{K^{0}} =D_{1\bar u}^{\bar K^{0}} = D_{1d}^{K^{+}} =D_{1\bar d}^{K^{-}}, \label{eq:DuK0} \\
&D_{1d}^{K^{0}} =D_{1\bar d}^{\bar K^{0}}= D_{1u}^{K^{+}}=D_{1\bar u}^{K^{-}}, \label{eq:DdK0}\\
&D_{1s}^{K^{0}} =D_{1\bar s}^{\bar K^{0}}= D_{1s}^{K^{+}}=D_{1\bar s}^{K^{-}}, \label{eq:DsK0}\\
&D_{1\bar u}^{K^{0}}=D_{1u}^{\bar K^{0}}= D_{1d}^{K^{-}}=D_{1\bar d}^{K^{+}}, \label{eq:DubarK0}\\
&D_{1\bar d}^{K^{0}} =D_{1d}^{\bar K^{0}} = D_{1u}^{K^{-}} =D_{1\bar u}^{K^{+}}, \label{eq:DdbarK0}\\
&D_{1\bar s}^{K^{0}} =D_{1s}^{\bar K^{0}}= D_{1s}^{K^{-}}=D_{1\bar s}^{K^{+}}.\label{eq:DsbarK0}
\end{align}
Here, for clarity, we omit arguments of fragmentation functions in Eqs.~(\ref{eq:DuK0}-\ref{eq:DsbarK0}).

For the unpolarized FF of $\rho$ meson, we take it similar to that of $K^*$ 
besides the strangeness suppression factor in the fragmentation process.
As usual, we differentiate between the favored and unfavored fragmentations. 
For the favored FF, we separate it into the leading and non-leading parts. 
The leading part is for hadron that contains the fragmenting quark and the non-leading part is the rest, 
i.e., we take, 
\begin{align}
&D_{1a}^{\rho,\rm favored} (z, \mu_0) = D_{1a}^{\rho,\rm favored,leading} (z, \mu_0)\nonumber\\ 
&\phantom{D_{1a}^{\rho,\rm favored} (z, \mu_0)}+D_{1a}^{\rho,\rm favored,nonleading} (z, \mu_0), \\
&D_{1a}^{\rho,\rm favored,nonleading} (z, \mu_0)= D_{1b}^{\rho,\rm unfavored} (z, \mu_0).
\end{align}

We relate those for $\rho$ to $K^*$ by,
\begin{align}
&D_{1\rm nonstrange}^{\rho,\rm favored,leading} (z, \mu_0) = D_{1\rm strange}^{K^{*},\rm favored,leading} (z, \mu_0), \\
&D_{1a}^{\rho,\rm unfavored} (z, \mu_0) = D_{1a}^{K^{*},\rm unfavored} (z, \mu_0)/\lambda_s, 
\end{align}
where $\lambda_s$ is the strangeness suppression factor and is simply taken as  $\lambda_s= 1/3$ 
in the numerical calculations presented in the following of this paper.
In this way, we obtain, e.g., 
\begin{align}
D_{1u}^{\rho^+} (z, \mu_0) &= D_{1\bar s}^{K^{*0}} (z, \mu_0)+\frac{1-\lambda_s}{\lambda_s} D_{1u}^{K^{*0}} (z, \mu_0), \\
D_{1d}^{\rho^+} (z, \mu_0) &= D_{1u}^{K^{*0}} (z, \mu_0)/\lambda_s,\\
D_{1u}^{\rho^0} (z, \mu_0) &= D_{1d}^{\rho^0} (z, \mu_0)\nonumber\\
&=\frac{1}{2}D_{1\bar s}^{K^{*0}} (z, \mu_0)+\frac{2-\lambda_s}{2\lambda_s} D_{1u}^{K^{*0}} (z, \mu_0), \\
D_{1s}^{\rho^+} (z, \mu_0) &=D_{1s}^{\rho^0} (z, \mu_0)=D_{1d}^{\rho^+} (z, \mu_0).
\end{align}

\subsection{The $S_{LL}$-dependent fragmentation function}
\label{sec:FFpol}

We take two different scenarios for the parameterizations of $S_{LL}$-dependent FFs. 
In the first scenario, we follow the same strategy employed in~\cite{Chen:2016iey} and differentiate 
between favored and unfavored fragmentations, i.e., 
\begin{align}
& D_{1LL}^{\rm unfavored} (z,\mu_0) = c_1 D_{1}^{\rm unfavored} (z,\mu_0), \label{eq:d1LLpar1a}\\
& D_{1LL}^{\rm favored} (z,\mu_0) = c_1 (a_1 z + 1) D_{1}^{\rm favored} (z,\mu_0),\label{eq:d1LLpar1b}
\end{align}
where $c_1$ and $a_1$ are two free parameters.  

In the second scenario, we adopt the same form of parameterizations for both favored and unfavored fragmentations. 
In this case, we find that the linear factor $az+1$ does not offer a good description to the data available~\cite{Ackerstaff:1997kj} 
and we change the power of $z$ to $1/2$, i.e., 
\begin{align}
D_{1LL} (z,\mu_0) = c_2 (a_2 z^{1/2} + 1) D_{1} (z,\mu_0), \label{eq:d1LLpar2}
\end{align} 
where $c_2$ and $a_2$ are free parameters.

From the condition that $-2\le D_{1LL}/D_{1}\le 1$, we obtain constraints for the parameters 
in the parameterizations given by Eqs.~(\ref{eq:d1LLpar1a}-\ref{eq:d1LLpar2}). 
They should be taken in the range $-2\le c_i\le1$ and $\min\{1/c_i,-2/c_i\}\le a_i+1\le \max\{1/c_i,-2/c_i\}$. 
It should be mentioned that by choosing the parameters in this region, 
we obtain the FF $D_{1LL}$ satisfying the positivity condition $-2\le D_{1LL}/D_{1}\le 1$ at the initial scale. 
It can be shown that, because the DGLAP evolution equation for $D_{1LL}$ is the same as that for $D_{1}$, 
the results obtained at other scales remain in this physical range. 
We also check this constantly in the numerical calculations.

We note that as we usually do for the spin dependent FFs,  
here we choose to parameterize the relationship between the $S_{LL}$-dependent FFs and unpolarized FFs in both scenarios. 
In this way, we take the standpoint that the unpolarized FFs are known to much higher accuracies than those for the polarized ones. 
However,  in reality, there are still very high uncertainties in the unpolarized FFs. 
We have also different parameterizations available. 
They can influence the values of the parameters in the parameterizations in both scenarios given 
by Eqs.~(\ref{eq:d1LLpar1a}-\ref{eq:d1LLpar2}) and also our numerical results for $pp\to VX$. 
In this paper, we choose two sets of parameterizations, AKK08~\cite{Albino:2008fy} and DHESS~\cite{deFlorian:2017lwf}, 
to show the influences. 

\subsection{Fits to the LEP data and results of $D_{1LL}$}
\label{sec:FFfit}

We fix the parameters in the parameterizations given by Eqs.~(\ref{eq:d1LLpar1a}-\ref{eq:d1LLpar2}) by applying a $\chi^2$ analysis with the data available~\cite{Abreu:1997wd,Ackerstaff:1997kj}.
For each value of parameters in Eqs.~(\ref{eq:d1LLpar1a}-\ref{eq:d1LLpar2}), we evolve the corresponding FFs utilizing 
the DGLAP equation given by Eq.~(\ref{eq:DGLAPD1LL}) to obtain a dataset of FFs at different factorization scales. 
Then we calculate the spin alignment of $K^{*0}$ with this dataset and compare the results obtained with the LEP data~\cite{Abreu:1997wd,Ackerstaff:1997kj} 
to get the corresponding $\chi^2$ value.   
We note that,  to be consistent with the LEP data~\cite{Abreu:1997wd,Ackerstaff:1997kj}, here as well as in the following of the paper, 
$K^{*0}$ represents the sum of $K^{*0}$ and its anti-particle $\bar K^{*0}$. 

\begin{figure}[htb]
  \begin{tabular}{cc}
    \includegraphics[width=0.21\textwidth]{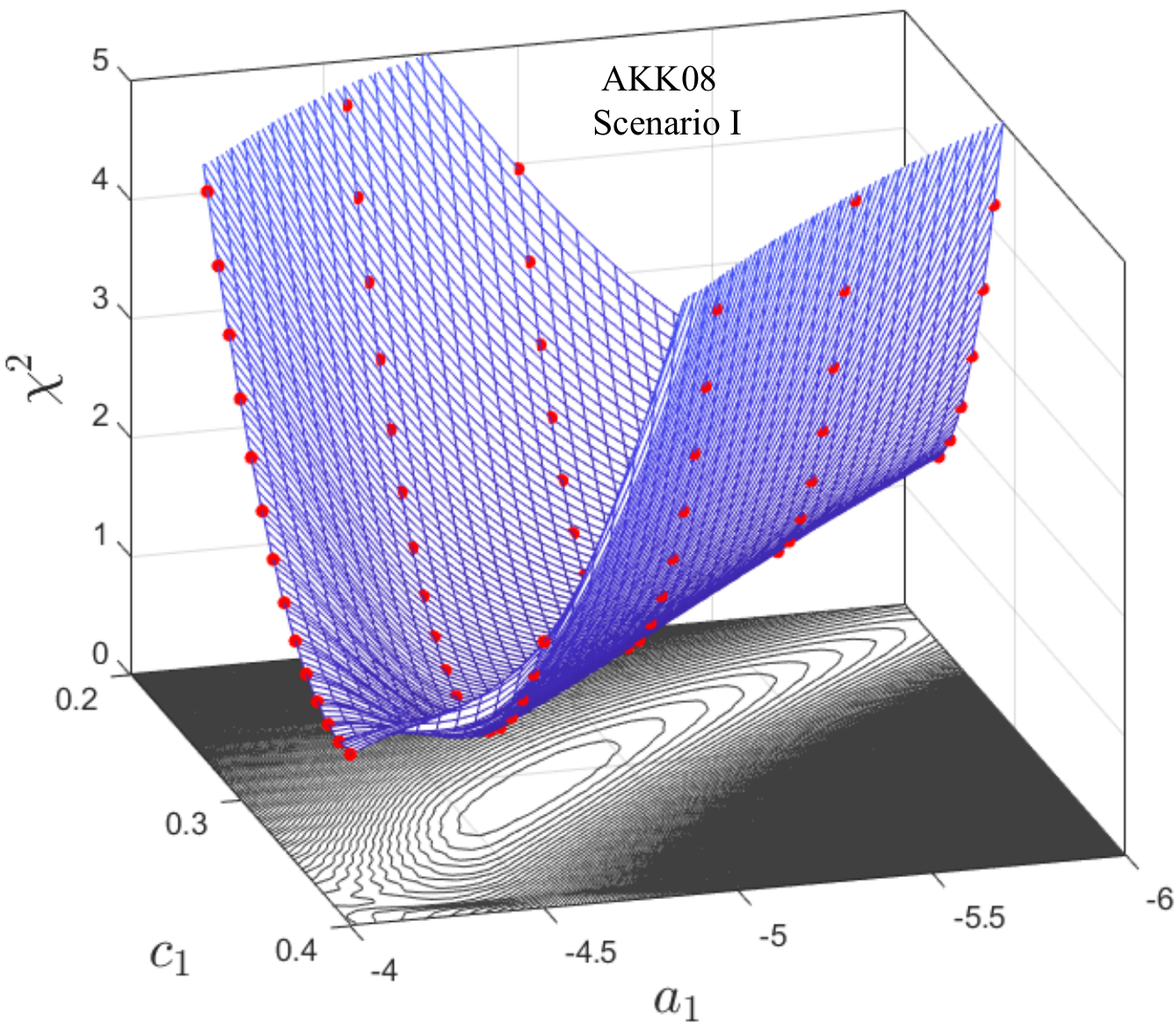}
    \includegraphics[width=0.25\textwidth]{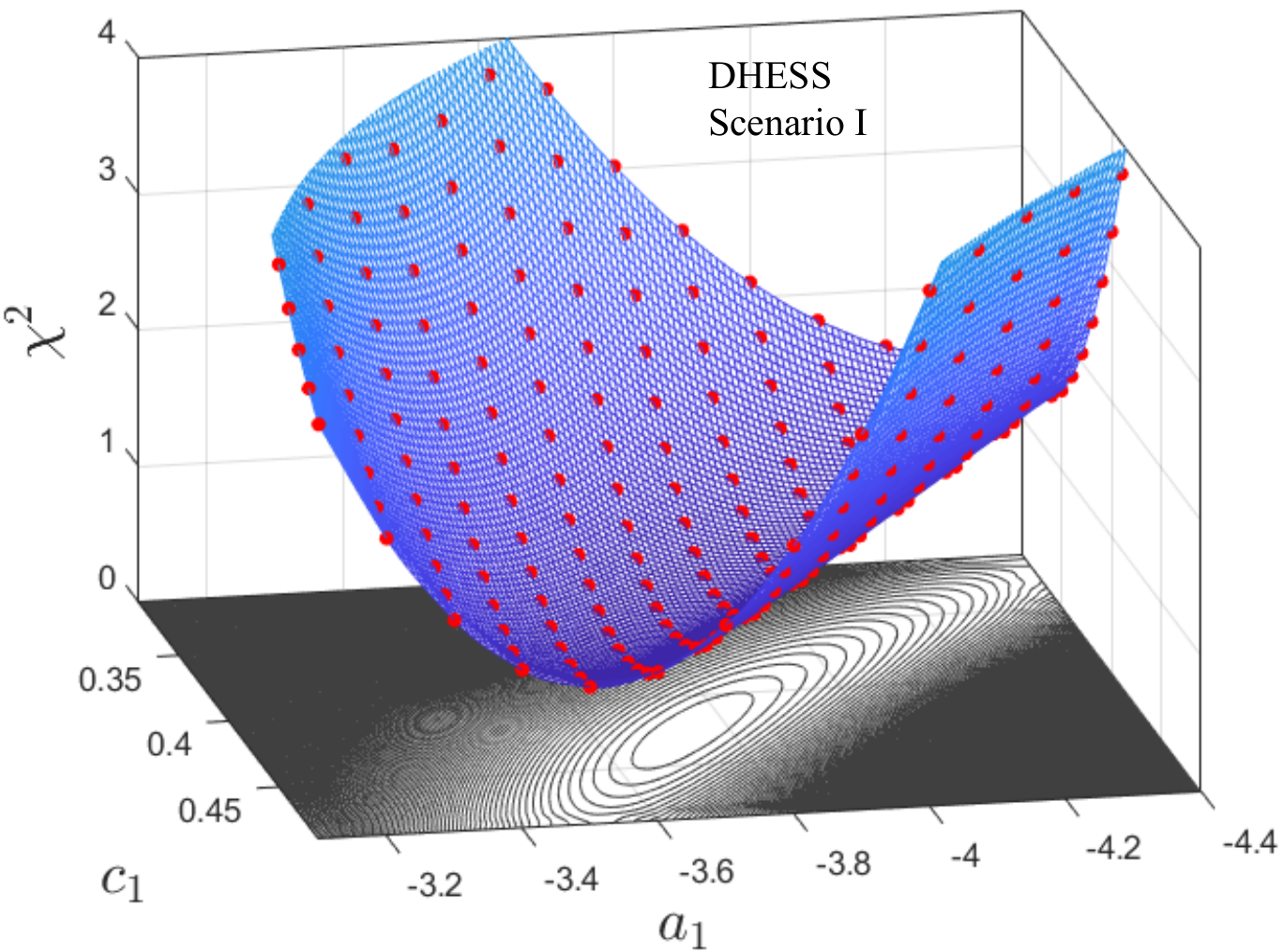}
  \end{tabular}
\caption{The $\chi^2$ plot in scenario I with AKK08 and DHESS parameterizations. 
Here, as well as in all figures in the following of this paper, $K^{*0}$ denotes the sum of $K^{*0}$ and $\bar K^{*0}$.}
\label{fig:chi2_akk08_s1}
\end{figure}

\begin{figure}[h!]
\includegraphics[width=0.475\textwidth]{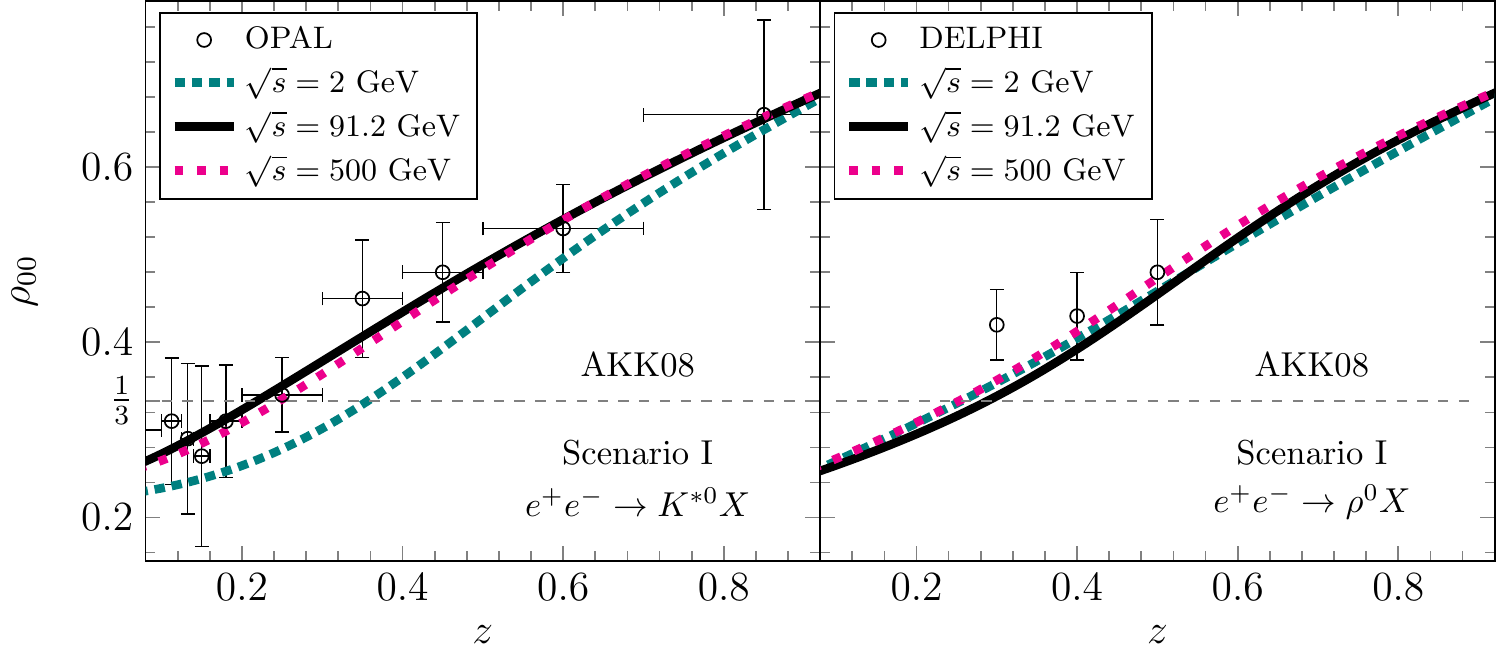}
\includegraphics[width=0.475\textwidth]{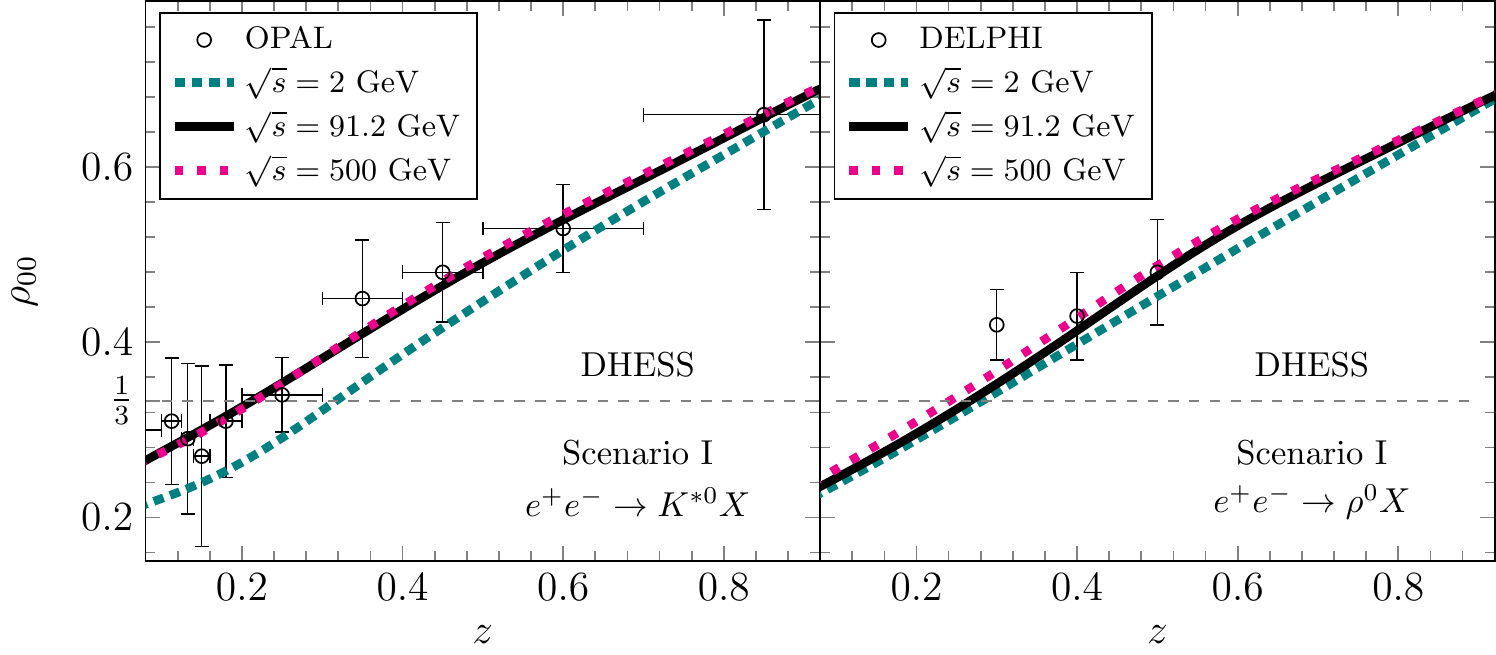}
\caption{(Color online) The spin alignments of $K^{*0}$ and $\rho^0$ in $e^+e^-\to VX$
at the $Z$-pole calculated in Scenario I with AKK08 and DHESS unpolarized FFs compared with experimental data~\cite{Abreu:1997wd,Ackerstaff:1997kj}. 
In the calculations, we have chosen the center-of-mass energy of $e^+e^-$ as the factorization scale.} 
\label{fig:fit-k-s1}
\end{figure}

Shown in Fig.~\ref{fig:chi2_akk08_s1} are the $\chi^2$ plots in scenario I with AKK08 and DHESS parameterizations for unpolarized FFs. 
The minimal values are $\chi^2=0.88$ at  $(c_1,a_1) = (0.24,-5.6)$ with AKK08 and $\chi^2=0.83$ at $(c_1,a_1)=(0.43,-3.7)$ with DHESS respectively. 
With these values of $c_1$ and $a_1$, we obtain the spin alignments of $K^{*0}$ and $\rho^0$ mesons 
and compare the results with data~\cite{Abreu:1997wd,Ackerstaff:1997kj} in Fig.~\ref{fig:fit-k-s1}. 
We also see that the data can be well described in both cases.    

From Fig.~\ref{fig:fit-k-s1}, we see that the scale dependence is more obvious in the small $z$ region but quite small at large $z$. 
It is also more obvious for $K^{*0}$ than that for $\rho^0$. 
To see where this difference comes from, we take AKK08 as an example to look at the corresponding results for FFs. 

\begin{figure}[h!]
\includegraphics[width=0.475\textwidth]{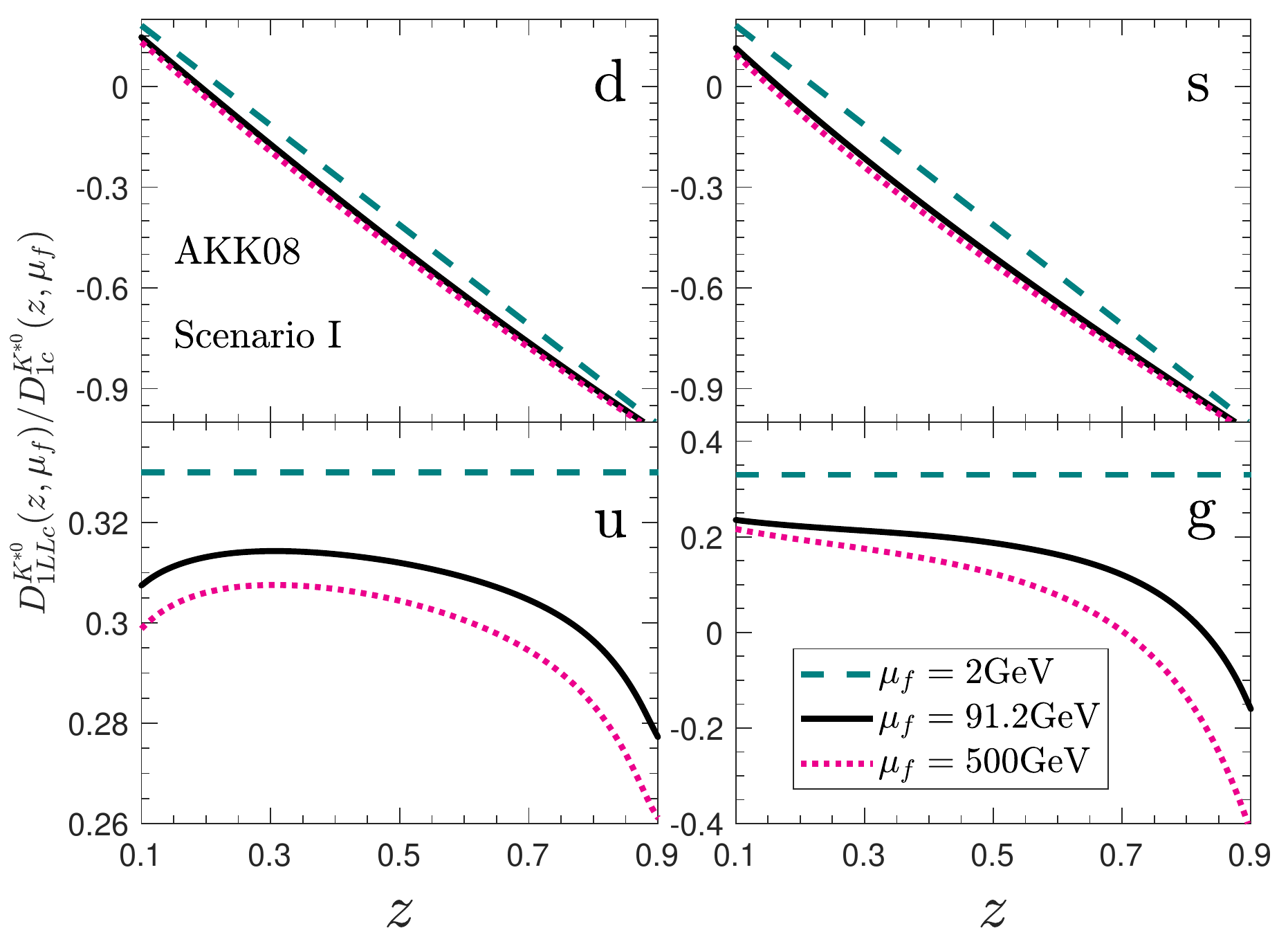}
\caption{(Color online) The ratio of the spin dependent fragmentation function $D_{1LL} (z, \mu_f)$ 
to that of the corresponding spin averaged $D_{1} (z,\mu_f)$ 
at different scales in Scenario I with AKK08 FFs.}
\label{fig:ratio-ffs-s1}
\end{figure}

\begin{figure}[h!]
\includegraphics[width=0.475\textwidth]{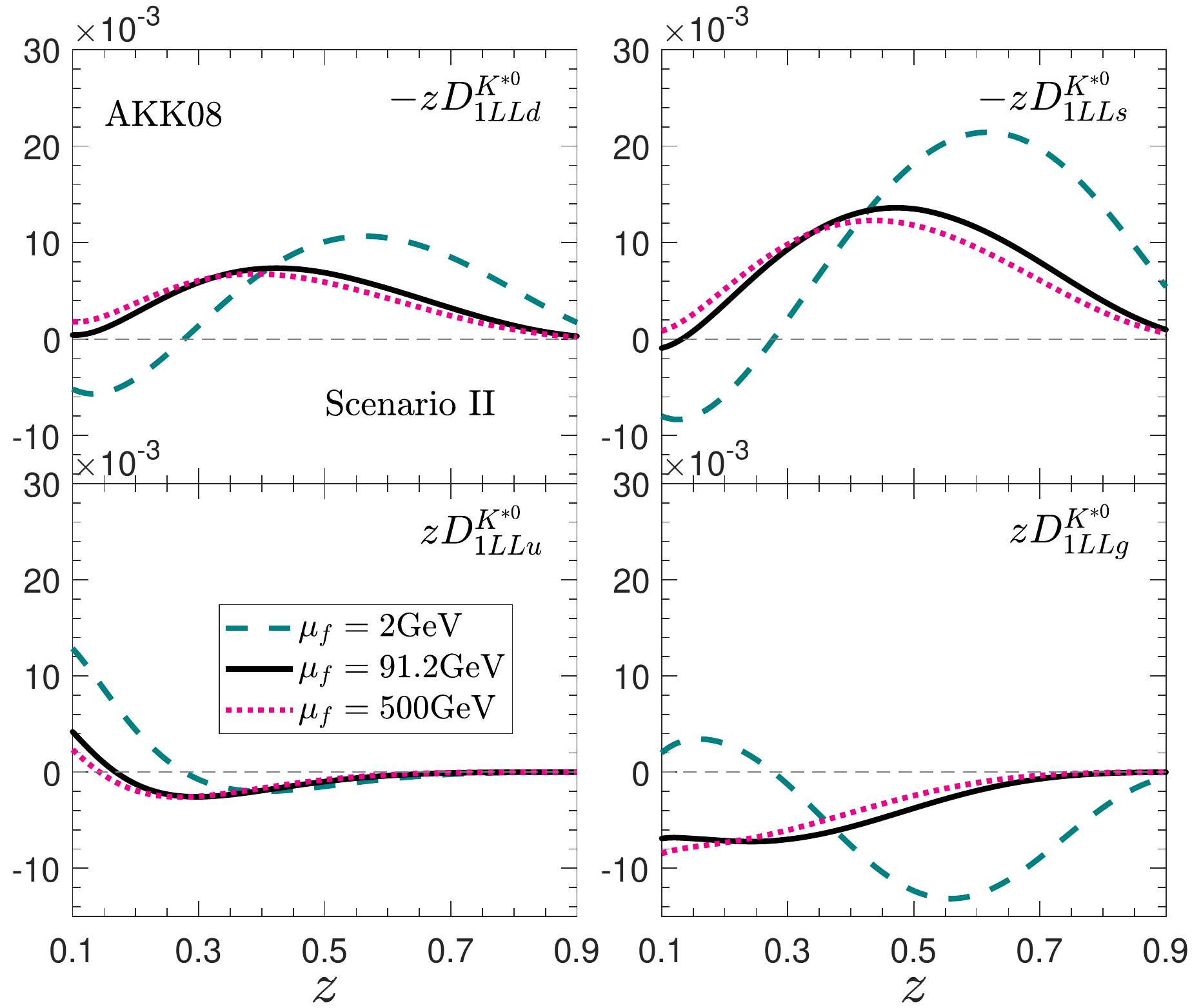}
\caption{(Color online) The spin dependent fragmentation function $D_{1LL} (z, \mu_f)$ at different scales in Scenario I with AKK08 FFs.}
\label{fig:D1LL-s1}
\end{figure}

In Fig.~\ref{fig:ratio-ffs-s1}, we show the ratios $D_{1LLc}^{K^{0*}}/D_{1c}^{K^{0*}}$ for different flavors of quarks and that of gluon with AKK08 FFs. 
The corresponding $S_{LL}$-dependent FFs $D_{1LLc}^{K^{0*}}$ are shown in Fig.~\ref{fig:D1LL-s1}, 

We note that for the production of $K^{*0}$, $u$-quark fragmentation is unfavored while $d$ and $s$ fragmentations are favored. 
From Fig.~\ref{fig:ratio-ffs-s1}, we see that, in scenario I, the ratio $D_{1LL}/D_1$ is almost the same for 
favored fragmentations of different flavors of quarks but it is very different from that for the unfavored quark fragmentation. 
It is negative and relatively larger in magnitude in most of the $z$ region in the favored case, 
but is positive and relatively smaller in the unfavored case. 
The scale dependence in the favored case is quite weak but seems much stronger in the unfavored case.
We see also that, though starting from the same ratio at the initial scale, the gluon fragmentation function 
behaves quite differently from the unfavored quark fragmentation function after the QCD evolution.
It becomes even negative at large $z$. 
This is because in QCD evolution to the first order, gluon splitting to a $q\bar q$-pair $g\to q\bar q$ and 
gluon radiation of a quark $q\to qg$ are considered. 
For the gluon fragmentation, after the gluon splitting $g\to q\bar q$, different flavors of quarks can be produced 
so that favored quark fragmentation can contribute thus brings large change to gluon FF. 
In contrast, for the unfavored quark fragmentation, after the gluon radiation of the quark $q\to qg$, 
the flavor of $q$ is unchanged and the fragmentation remains unfavored.  

From Fig.~\ref{fig:D1LL-s1}, we see similar behaviors as those for the corresponding ratios in Fig.~\ref{fig:ratio-ffs-s1}. 
We see again similar behaviors for the favored FFs that are very different from the unfavored FF 
and also different from gluon FF. 
Here, we see explicitly that favored FFs dominate at larger $z$ while unfavored and gluon fragmentations play important roles at small $z$. 
We also see that because of the strangeness suppression in fragmentation, the leading contributions 
from $s$-quark fragmentation is much larger than that from $d$-quark. 

From the results shown in Figs.~\ref{fig:ratio-ffs-s1} and \ref{fig:D1LL-s1}, 
we can now understand why there is a slight difference between the scale dependence 
of the spin alignment of $K^{*0}$ and $\rho^0$ as shown in Fig.~\ref{fig:fit-k-s1}. 
Because of the strangeness suppression in the favored $d$-fragmentation, 
contributions from unfavored quark and gluon fragmentations are relatively larger 
for the production of $K^{*0}$ than that of $\rho^0$. 
The stronger scale dependence of $D_{1LL}/D_1$ for the unfavored and gluon fragmentation 
leads to a slightly stronger scale dependence of the spin alignment of $K^{*0}$ than that of $\rho^0$.

The corresponding results of scale dependence with DHESS FFs are similar, and we skip the detailed discussions here .  

\begin{figure}[htb]
  \begin{tabular}{cc}
    \includegraphics[width=0.22\textwidth]{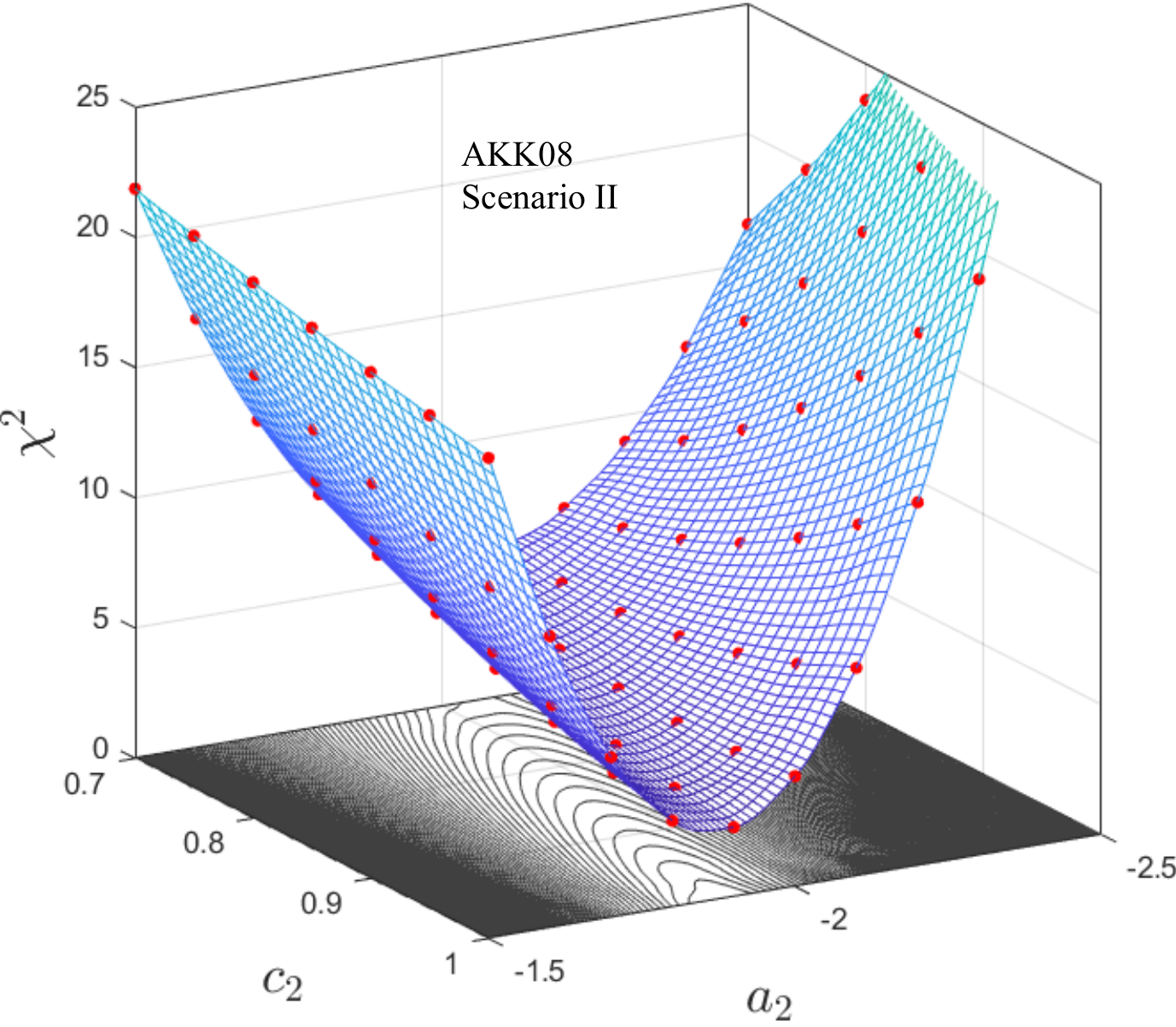}
    \includegraphics[width=0.23\textwidth]{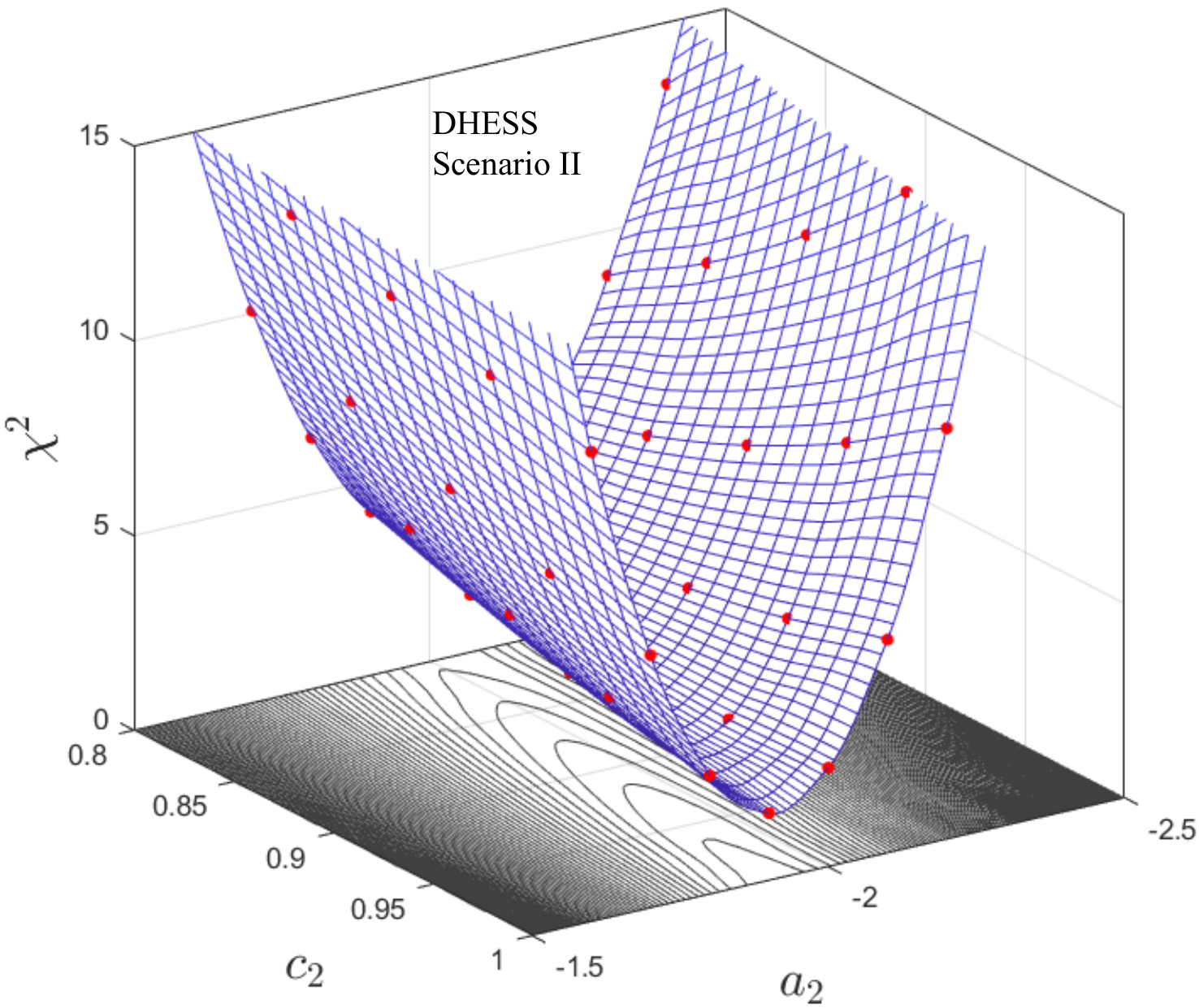}
  \end{tabular}
\caption{The $\chi^2$ plot in scenario II with AKK08 and DHESS parameterizations.}
\label{fig:chi2_akk08_s2}
\end{figure}

The calculation in the second scenario is similar. 
The obtained $\chi^2$ distributions are shown in Fig.~\ref{fig:chi2_akk08_s2}. 
Here, we see that with this scenario for polarized FFs, the $\chi^2$ distributions in the case with AKK08 unpolarized FFs 
and that with DHESS are quite similar to each other.   
The minimal $\chi^2$ value flows in the valley where $a_2 \sim -1.9$.  
With AKK08, we reach the minimal $\chi^2=2.68$ in physical regions of $c_2$ and $a_2$ at $(c_2,a_2)=(1.0, -1.9)$ 
and $\chi^2=1.78$ at the same values of $c_2$ and $a_2$ with DHESS. 

\begin{figure}[h!]
\includegraphics[width=0.475\textwidth]{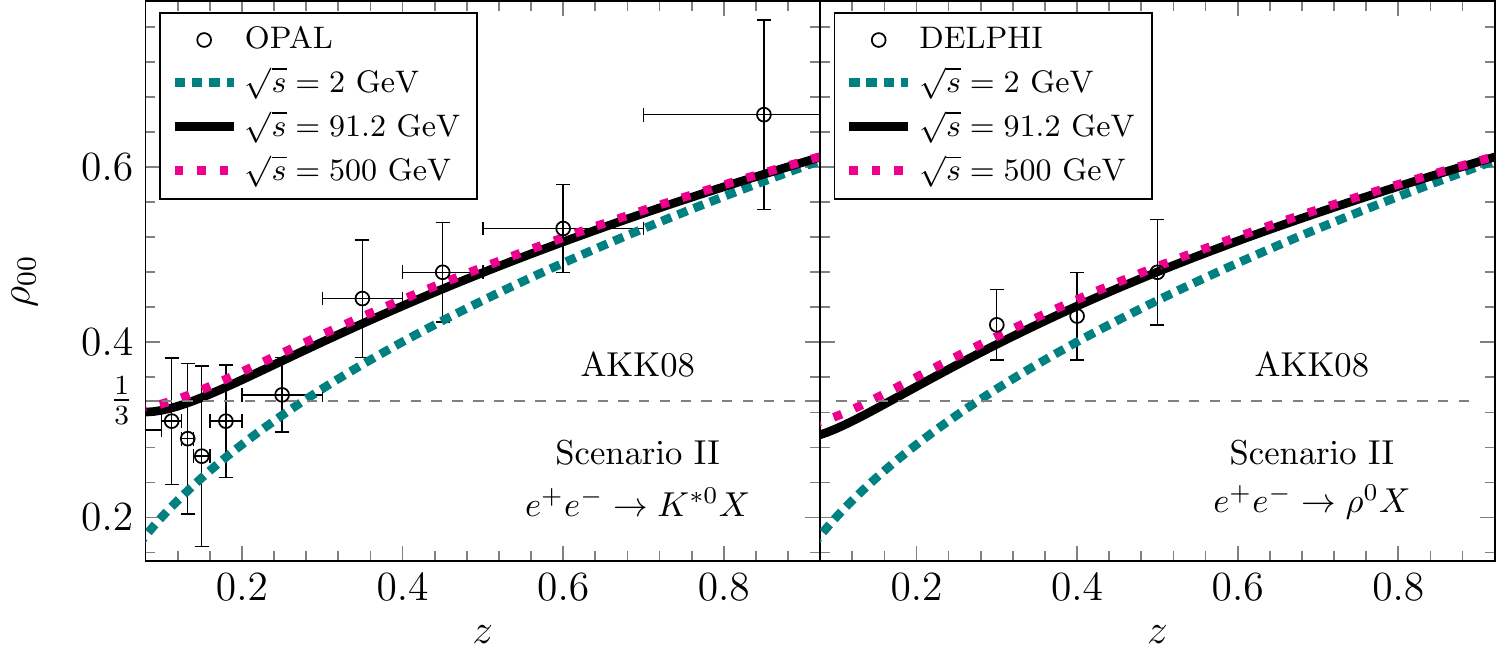}
\includegraphics[width=0.475\textwidth]{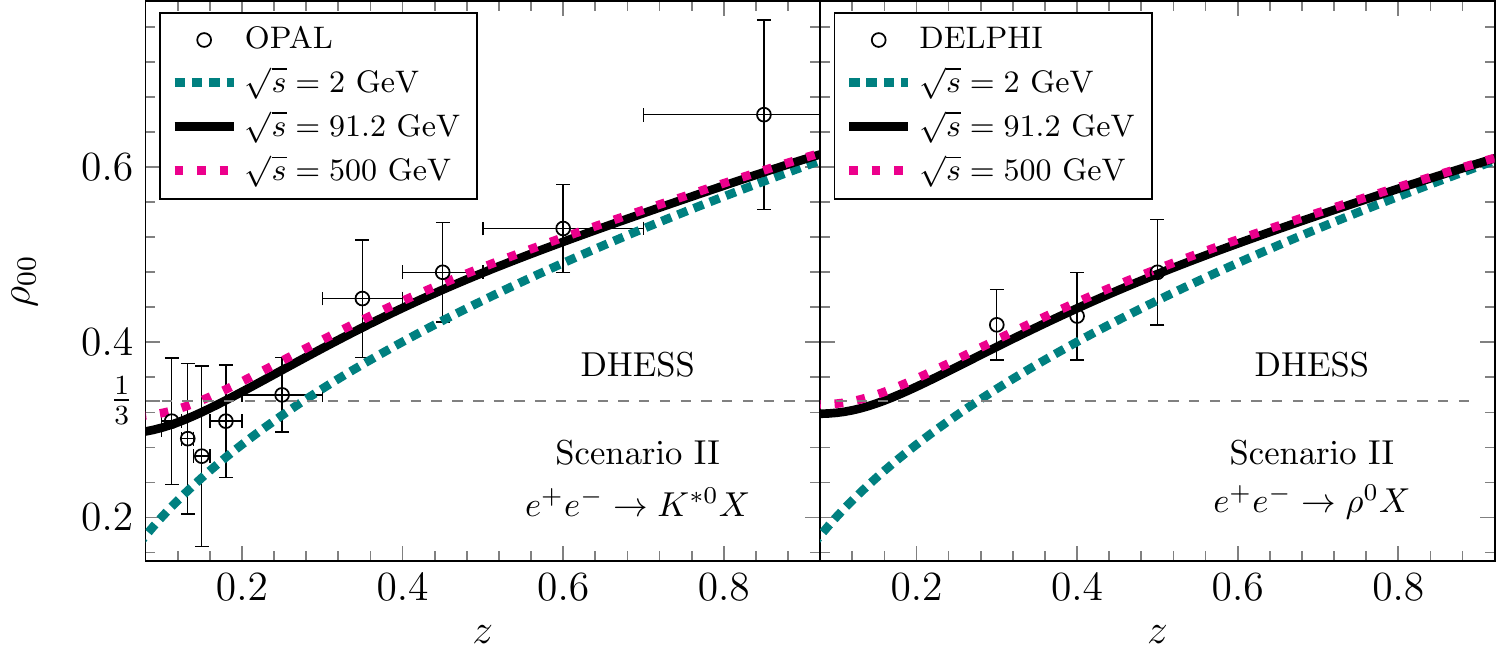}
\caption{(Color online) The spin alignment of $K^{*0}$ (left panel) and that of $\rho^0$ (right panel) in $e^+e^-\to VX$
at the $Z$-pole calculated in Scenario II with AKK08 and DHESS FFs compared with experimental data~\cite{Abreu:1997wd,Ackerstaff:1997kj}. }
\label{fig:fit-k-s2}
\end{figure}

We then calculate the spin alignment of $K^{*0}$ and $\rho^{0}$ mesons with these parameters 
and compare with data in Fig.~\ref{fig:fit-k-s2}. 
We see that a reasonable agreement with the data can also be achieved in this case. 

\begin{figure}[h!]
\includegraphics[width=0.475\textwidth]{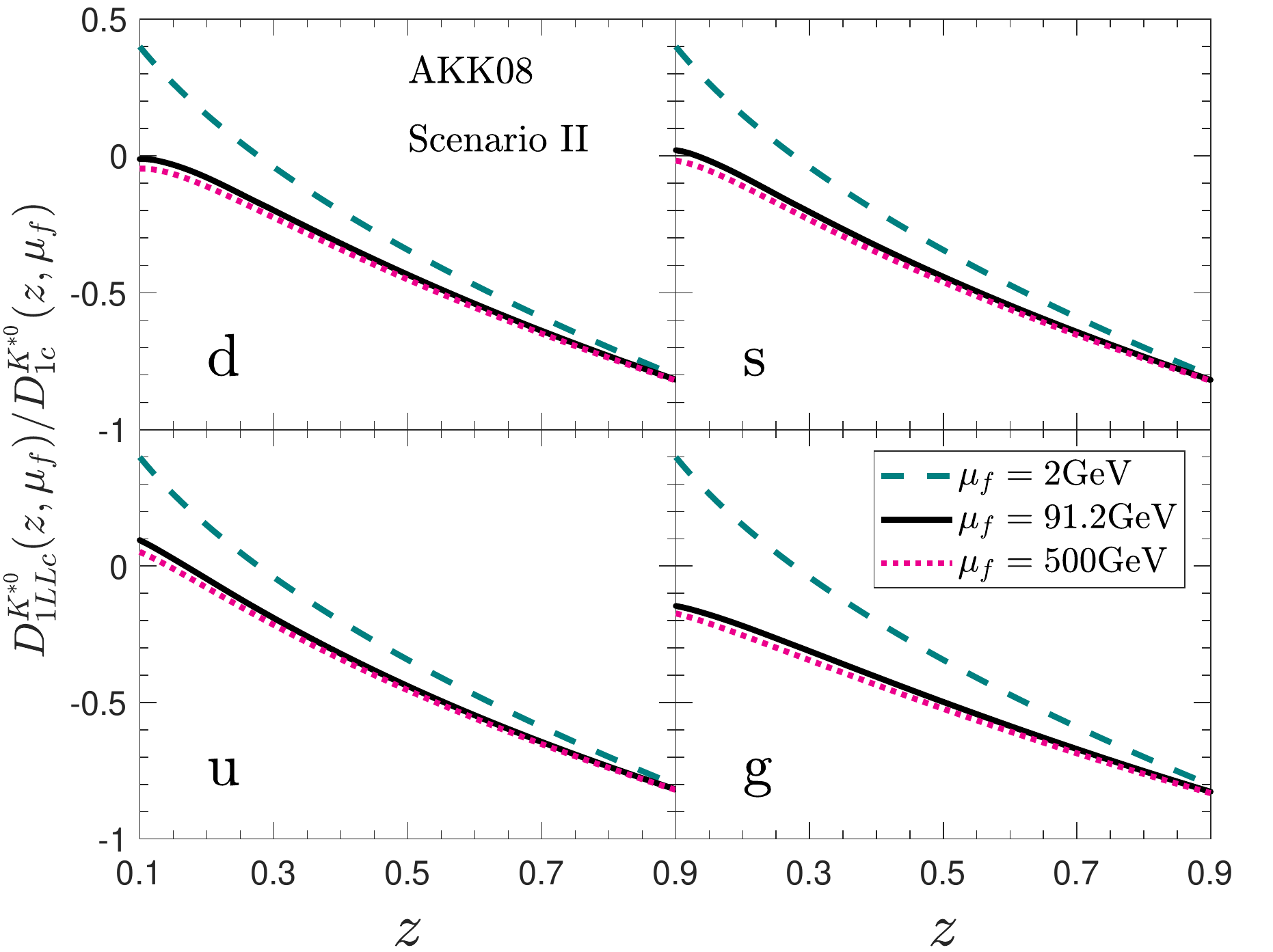}
\caption{(Color online) The ratio of the spin dependent fragmentation function $D_{1LL} (z, \mu_f)$ 
to that of the corresponding spin averaged $D_{1} (z,\mu_f)$ at different scales in Scenario II with AKK08 FFs.}
\label{fig:ratio-ffs-s2}
\end{figure}

\begin{figure}[h!]
\includegraphics[width=0.475\textwidth]{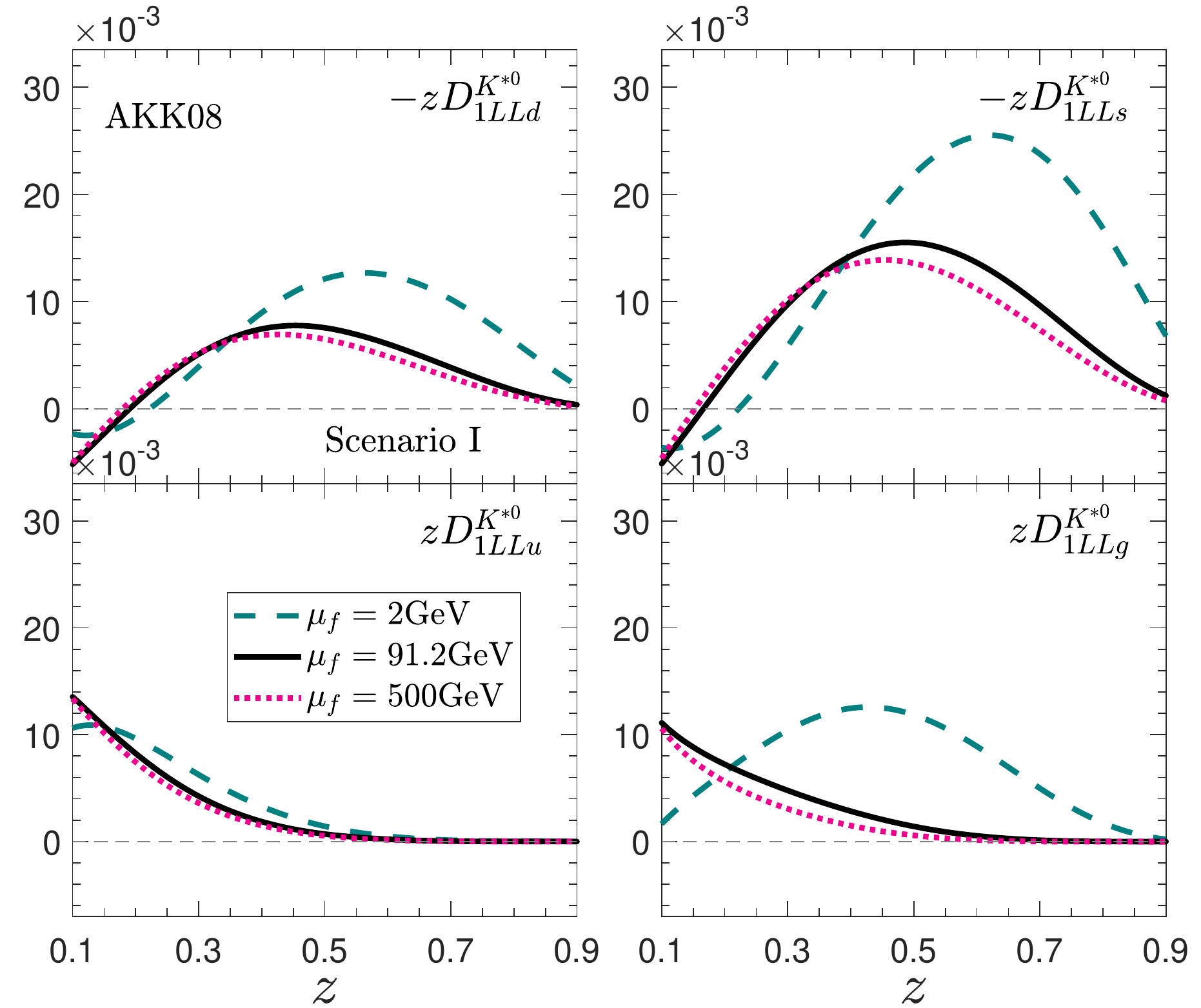}
\caption{(Color online) The spin-dependent fragmentation function $D_{1LL} (z, \mu_f)$ at different scales in Scenario II with AKK08 FFs.}
\label{fig:D1LL-s2}
\end{figure}

We again take AKK08 as an example and show our results obtained in this scenario 
of the ratios $D_{1LLc}^{K^{0*}}/D_{1c}^{K^{0*}}$ and the corresponding $S_{LL}$-dependent FFs $D_{1LLc}^{K^{0*}}$ 
in Figs.~\ref{fig:ratio-ffs-s2} and \ref{fig:D1LL-s2} respectively.

From Fig.~\ref{fig:ratio-ffs-s2}, we see that the ratios $D_{1LL}/D_1$ in this scenario for favored, unfavored and gluon fragmentations 
are quite similar with each other. 
By starting with the same parameterization at the initial scale, we obtain similar results after the QCD evolution.
The tiny differences are resulted from the differences in the corresponding unpolarized FFs.
Also because there is no large difference in the ratios $D_{1LL}/D_1$ 
between the favored and unfavored fragmentations in this scenario, 
we do not see similar difference in Fig.~\ref{fig:fit-k-s2} between the spin alignment of $K^{*0}$ 
and that of $\rho^0$ in this scenario as that shown in Fig.~\ref{fig:fit-k-s1} in scenario I.  

Because of the differences in the corresponding unpolarized FFs, 
the obtained $D_{1LL} (z, \mu_f)$ shown in Fig.~\ref{fig:D1LL-s2} 
exhibits also quite large differences between the favored and unfavored quark fragmentation and that of gluon.  
Here we see that, similar to those in scenario I, the favored FFs also dominate at larger $z$ but the unfavored 
and gluon FFs may have large contribution in the small $z$-region. 
The gluon FF $D_{1LLg} (z, \mu_f)$ is negative and quite large in magnitude for small $z$ and 
should play an important role in this region.

Comparing the FFs obtained in the two different scenarios, we see quite large differences. 
Nevertheless the obtained spin alignments in both cases can describe the LEP data~\cite{Abreu:1997wd,Ackerstaff:1997kj}.  
This is because the freedom to choose different parameterizations is quite large, 
the LEP data~\cite{Abreu:1997wd,Ackerstaff:1997kj} alone can not fix them to high accuracy.  
In this connection, we note that we have not considered the flavor dependence of the ratio between the unpolarized 
and the $S_{LL}$-dependent FFs besides different choices for the favored and unfavored fragmentation in scenario I.  
It is clear that more data in different reactions are necessary in order to determine these FFs to high precisions.

\section{Numerical results for $pp\to VX$}
\label{sec:Numpp}

In this section, we apply the FFs obtained in Sec.~\ref{sec:FF} to $pp\to VX$ and calculate the spin alignment of vector mesons numerically. 
To have a better understanding of the results in such a complicated process, 
we first present the fractional production rate of different flavor of partons. 
After that, we show our predictions on the spin alignment of $K^{*0}$ and $\rho^0$ mesons in both scenarios.
We recall that, for all the results presented throughout the paper, to be consistent with the LEP data~\cite{Abreu:1997wd,Ackerstaff:1997kj}, 
$K^{*0}$ represents the sum of $K^{*0}$ and its anti-particle $\bar K^{*0}$.

\subsection{Contributions of different flavors}
\label{sec:Numpp1}

From Eq.~(\ref{eq:x-pp-vX}), we can calculate contributions from different subprocesses to the cross section separately. 
The fractional contribution from a given type of parton $c$ to jet production is given by,
\begin{align}
R_c^{\rm jet}(y,p_{T}) =& \frac{d\sigma_{pp\to cX}}{dyd^2p_{T}}\Big/\sum_c\frac{d\sigma_{pp\to cX}}{dyd^2p_{T}},\label{eq:rJet}\\
\frac{d\sigma_{pp\to cX}}{dyd^2p_{T}}=&\sum_{abd} \int d y_2  x_1 f_a (x_1,\mu_f)  \nonumber\\
&\times x_2 f_b(x_2,\mu_f) \frac{1}{\pi} \frac{d\hat\sigma_{ab\to cd}}{d\hat t}. 
\label{eq:xJet}
\end{align}
Similarly, the fractional contribution to vector meson production is given by,
\begin{align}
R_c^{\rm V}(y,p_{T}) =& \frac{d\sigma_{pp\to cX\to VX}}{dyd^2p_{T}}\Big/\frac{d\sigma_{pp\to VX}}{dyd^2p_{T}},\label{eq:rVp}\\
\frac{d\sigma_{pp\to cX\to VX}}{dyd^2p_{T}} =& \sum_{abd} \int d y_2 \int \frac{dz}{z^2} x_1 f_a (x_1,\mu_f) \nonumber\\
&\times x_2 f_b(x_2,\mu_f) \frac{1}{\pi} \frac{d\hat\sigma_{ab\to cd}}{d\hat t}D_{1c}^V(z,\mu_f).
\label{eq:xVp}
\end{align}

\begin{figure}[h!]
\includegraphics[width=0.475\textwidth]{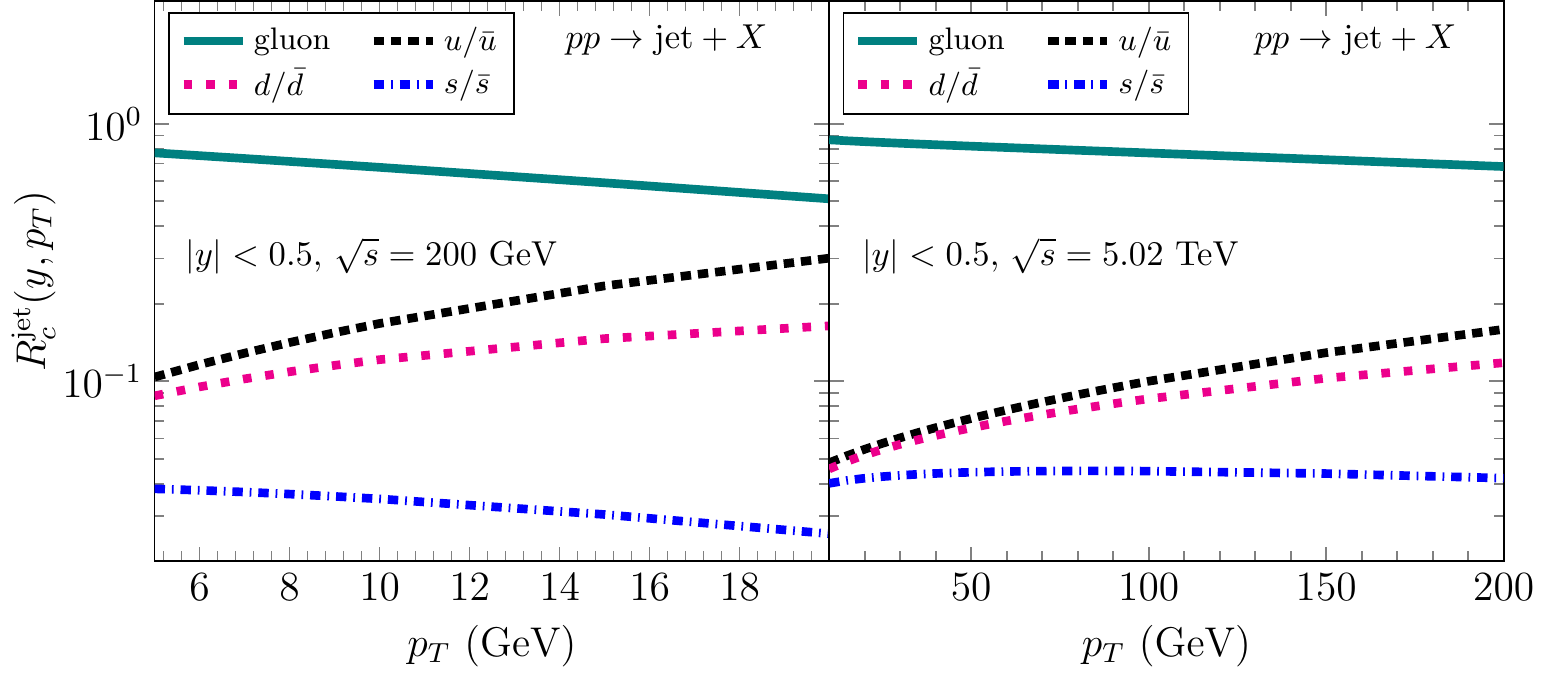}
\caption{(Color online) Fractional contributions $R_c^{\rm jet}(y,p_{T})$ to jet production from different flavors of quarks/anti-quarks 
and gluon at $|y|<0.5$ as functions of $p_{T}$ in $pp$ collisions at RHIC energy $\sqrt{s}=200$ GeV (left) and LHC energy $\sqrt{s}=5.02$ TeV (right).}
\label{fig:rJet1}
\end{figure}

\begin{figure}[h!]
\includegraphics[width=0.475\textwidth]{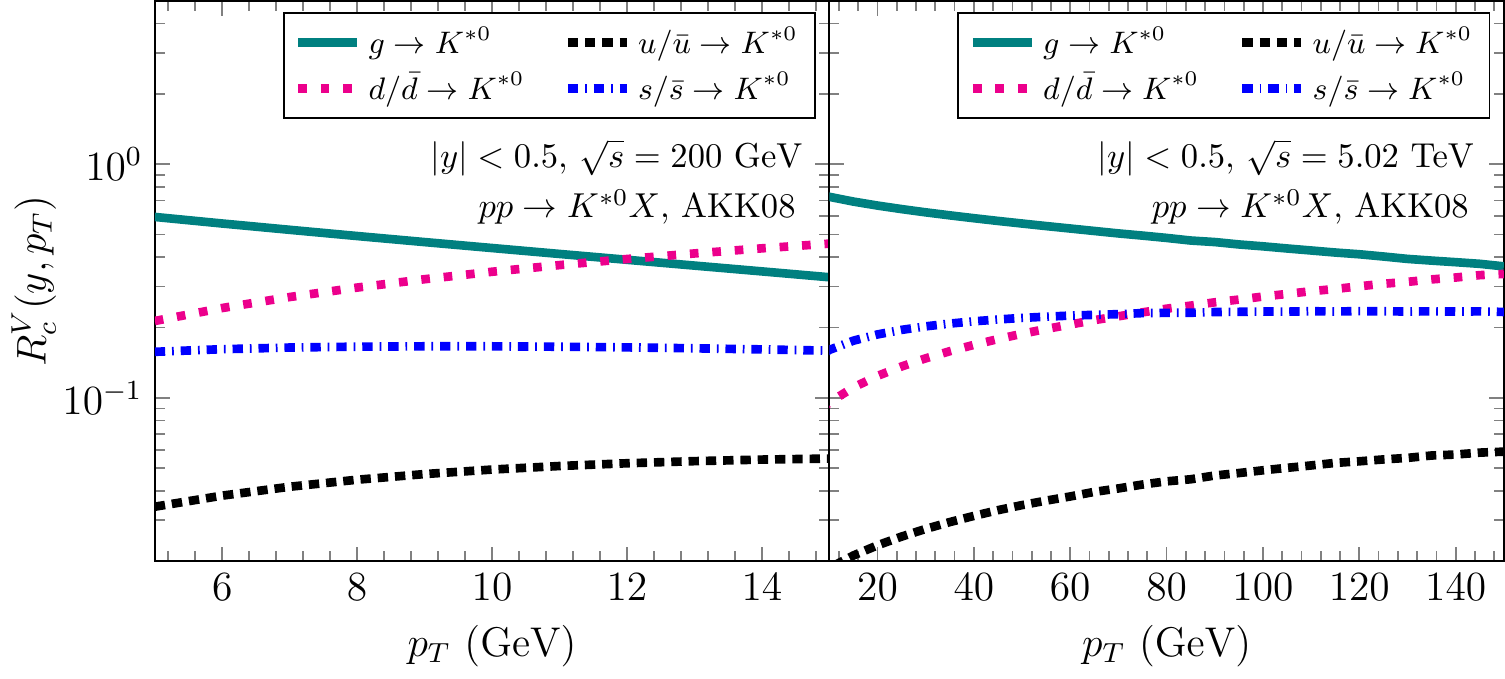}
\includegraphics[width=0.475\textwidth]{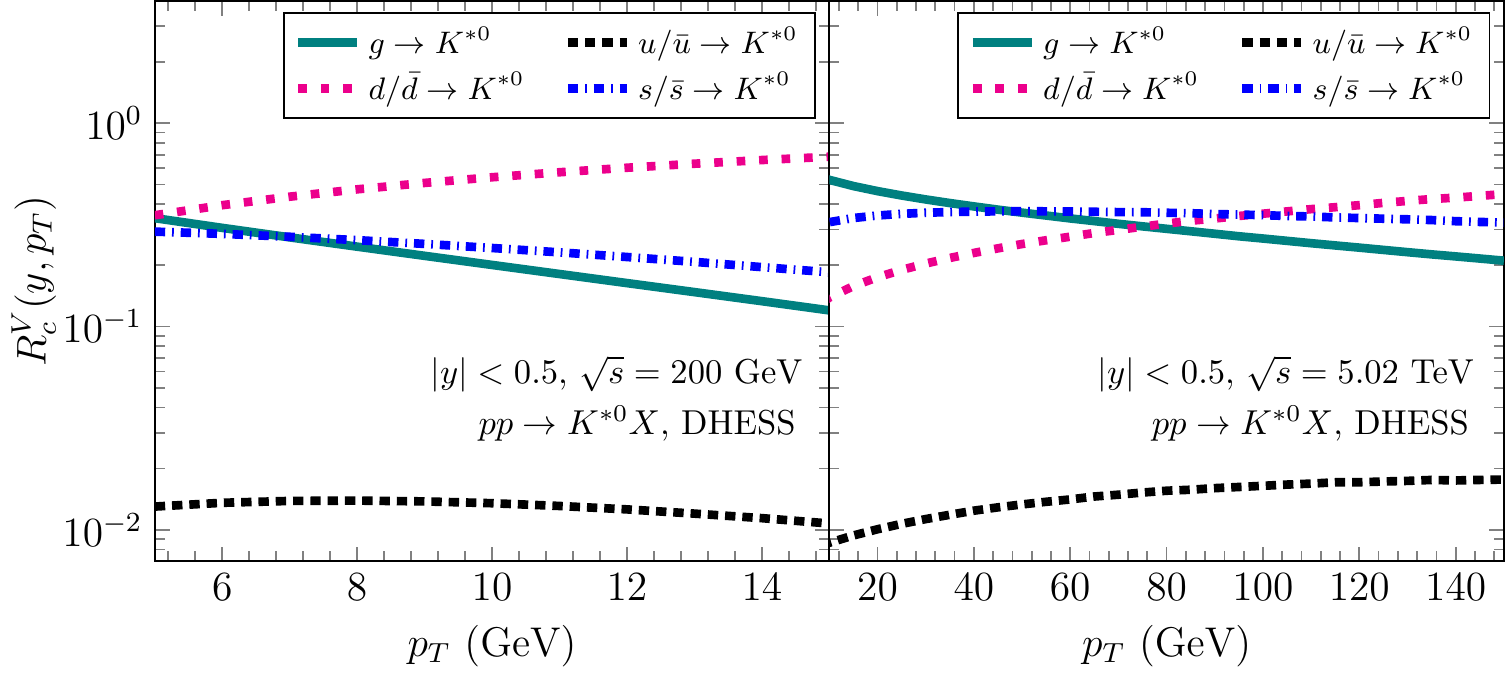}
\caption{(Color online) Fractional contributions $R_c^{\rm V}(y,p_{T})$ to the production of $K^{*0}$ 
from different flavors of quarks/anti-quarks and gluon at $|y|<0.5$ as functions of $p_{T}$ 
in $pp$ collisions at RHIC energy $\sqrt{s}=200$ GeV (left) and LHC energy $\sqrt{s}=5.02$ TeV (right)
obtained with AKK08 and DHESS FFs respectively.}
\label{fig:rVp1}
\end{figure}

In Fig.~\ref{fig:rJet1}, we show the results of $R_c^{\rm jet}(y,p_{Tc})$ calculated from Eqs.~(\ref{eq:rJet}) and (\ref{eq:xJet}) at the RHIC and LHC energies in the middle rapidity as functions of $p_T$. 
Taking $K^{*0}$ as an example, we show the corresponding results of $R_c^{\rm V}(y,p_{T})$ 
calculated with Eqs.~(\ref{eq:rVp}) and (\ref{eq:xVp}) in Fig.~\ref{fig:rVp1}. 


From Fig.~\ref{fig:rJet1}, we see that in the presented $p_T$ regions, 
the gluon contribution dominates at both RHIC and the LHC energies for jet productions. 
The $u/\bar u$ contribution is the largest among the three flavors of quarks while $s/\bar s$ is the smallest.  
This results from the differences in PDFs~\cite{Dulat:2015mca} for different flavors of partons. 

However, when FFs are taken into account, from Fig.~\ref{fig:rVp1}, we see that the gluon contribution becomes less dominate. 
With AKK08 FFs the $d/\bar d$ contribution is even larger than the gluon contribution at the RHIC energy for $p_T>12$ GeV 
while $u/\bar u$ contribution becomes the smallest one.
With DHESS FFs the $d/\bar d$ and $s/\bar s$ contributions dominate at almost all $p_T$ range at the RHIC energy, 
and they dominate at the LHC energy for $p_T>50$ GeV. 
The exact value of $R_c^V$ depends on the specific FF parameterizations, but the overall trends are similar.
This is because the differential cross section for the production of parton $c$ decreases 
very fast with increasing $p_T$, much faster than the FF of $c\to VX$ decreases with increasing $z$.
Usually the $z$-dependence of FF is much smoother compared with the $p_T$-dependence of the cross section.
As a result, in the large $p_T$ region for hadron production, contributions from relatively large $z$ (say $z>0.3$) dominate.  
The leading contribution from favored quark fragmentations plays a more and more important role with increasing $p_T$.  

From Fig.~\ref{fig:rVp1}, we also see that, by studying the $p_T$ dependence in the central rapidity region, 
we can study the interplay of contributions from gluon and favored quark fragmentation,  
while at the LHC, we mainly study the contribution from gluon fragmentation.
Quark fragmentations should dominate the fragmentation regions in the collision processes, i.e., at very forward or very backward rapidities. 

\begin{figure}[h!]
\includegraphics[width=0.475\textwidth]{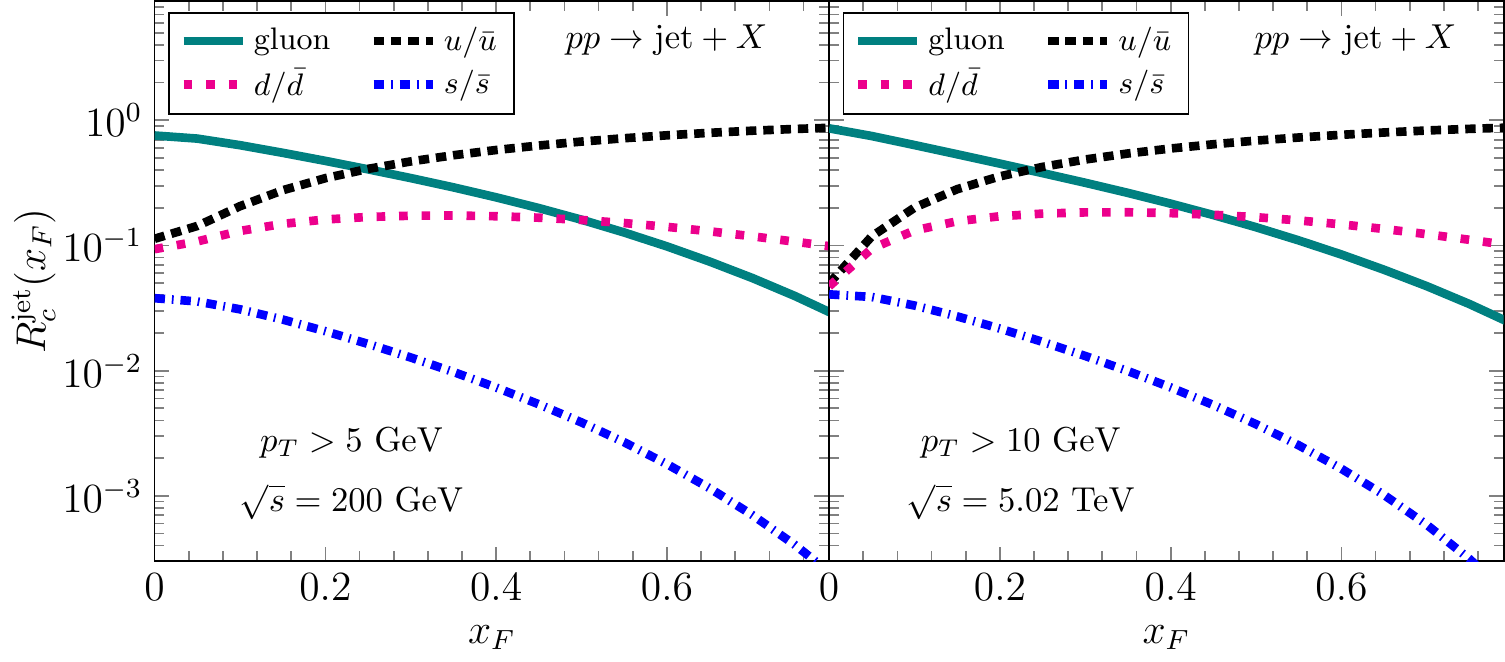}
\caption{(Color online) Fractional contributions $R_c^{\rm jet}(y,p_{T})$ to jet production from different flavors of 
quarks/antiquarks and gluon as functions of $x_F$ in $pp$ collisions at RHIC energy $\sqrt{s}=200$ GeV with $p_{T}\ge 5$ GeV (left)  
and LHC energy $\sqrt{s}=5.02$ TeV $p_{T}\ge 10$ GeV (right). }
\label{fig:rJet2}
\end{figure}

\begin{figure}[h!]
\includegraphics[width=0.475\textwidth]{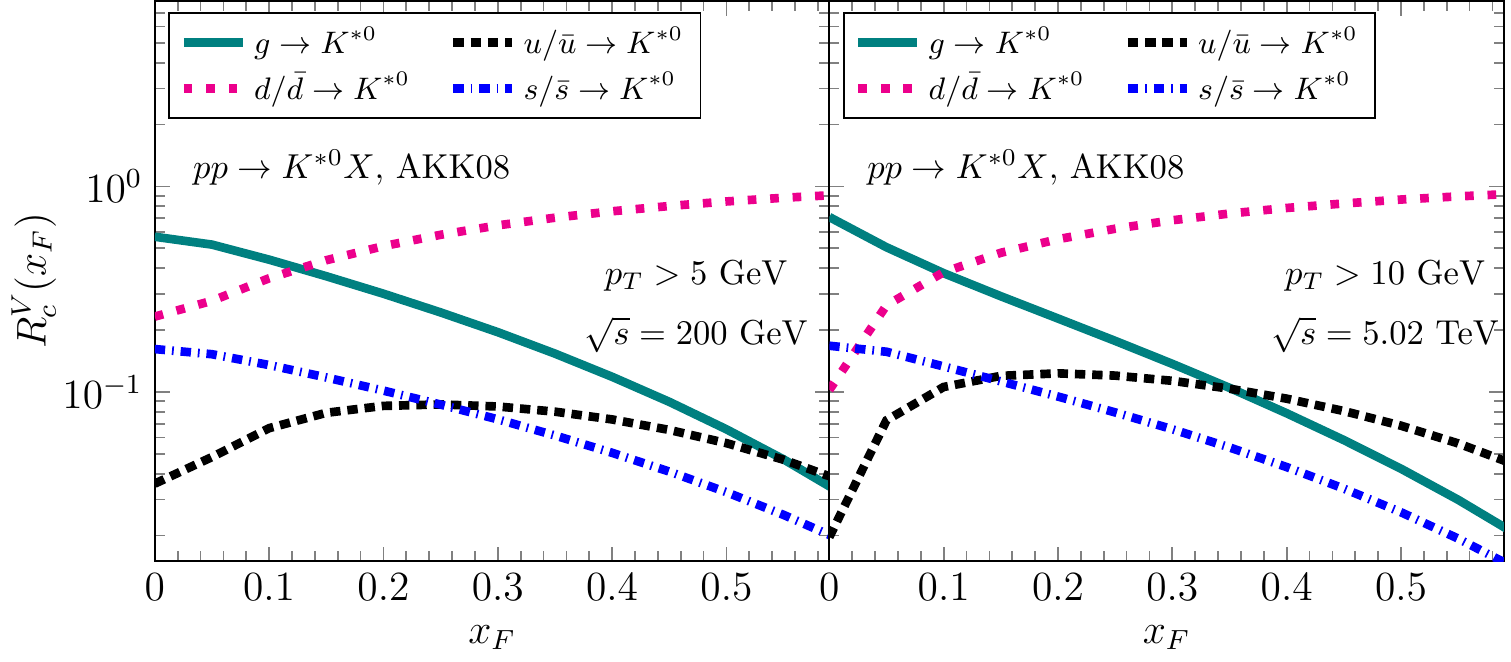}
\includegraphics[width=0.475\textwidth]{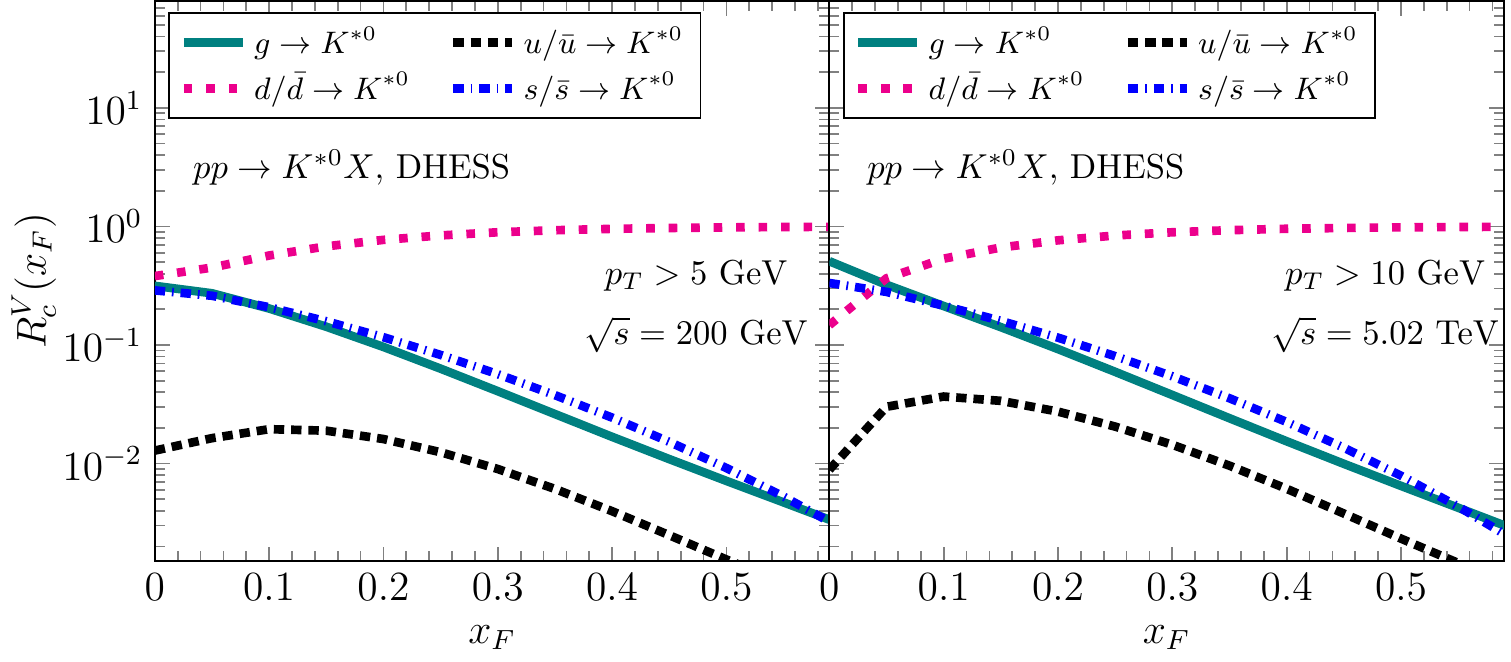}
\caption{(Color online) Fractional contributions $R_c^{\rm V}(y,p_{T})$ to the production of $K^{*0}$ from different flavors of quarks/antiquarks 
and gluon as functions of $x_F$ in $pp$ collisions at RHIC energy $\sqrt{s}=200$ GeV with $p_{T}\ge 5$ GeV (left)  
and LHC energy $\sqrt{s}=5.02$ TeV with $p_{T}\ge 10$ GeV (right) 
obtained with AKK08 and DHESS FFs respectively.}
\label{fig:rVp2}
\end{figure}

To see the behaviors at the fragmentation regions explicitly, in Figs.~\ref{fig:rJet2} and \ref{fig:rVp2}, 
we show the corresponding results at the RHIC energy with $p_T>5$ GeV and those 
at the LHC energy with $p_T>10$ GeV as functions of $x_F$.

From Fig.~\ref{fig:rJet2} and \ref{fig:rVp2}, we see clearly that in the large $x_F$ region quark contribution dominates.  
For jet production, $u/\bar u$ plays the most important role. 
Taking the FFs into account, for $K^{*0}$-production, the favored fragmentation from $d/\bar d$ dominates.
Hence, by studying hadron production at larger $x_F$, we study predominately the favored quark fragmentation.  

At the end of this part, we emphasize that, by studying vector meson production in $pp\to VX$ for large $p_T$ 
at RHIC and LHC energies, even in the central rapidity regions, contributions from FFs at relatively large $z$ dominate.  
From the results for FFs obtained in Sec.~\ref{sec:FFpol}, we find $D_{1LL}$ is significantly different from zero as well in the relatively large $z$ region. 
This leads us to the expectation that the vector meson spin alignment should be quite significant in $pp$ collisions.

\subsection{The spin alignment in $pp\to VX$}
\label{sec:Numpp2}

Using the spin-dependent FFs obtained in Sec.~\ref{sec:FFpol},
we calculate the spin alignment of vector meson in $pp\to VX$ using Eqs.~(\ref{eq:rho00}) and (\ref{eq:csSLL}). We present our predictions in the following.

\begin{figure}[h!]
\includegraphics[width=0.475\textwidth]{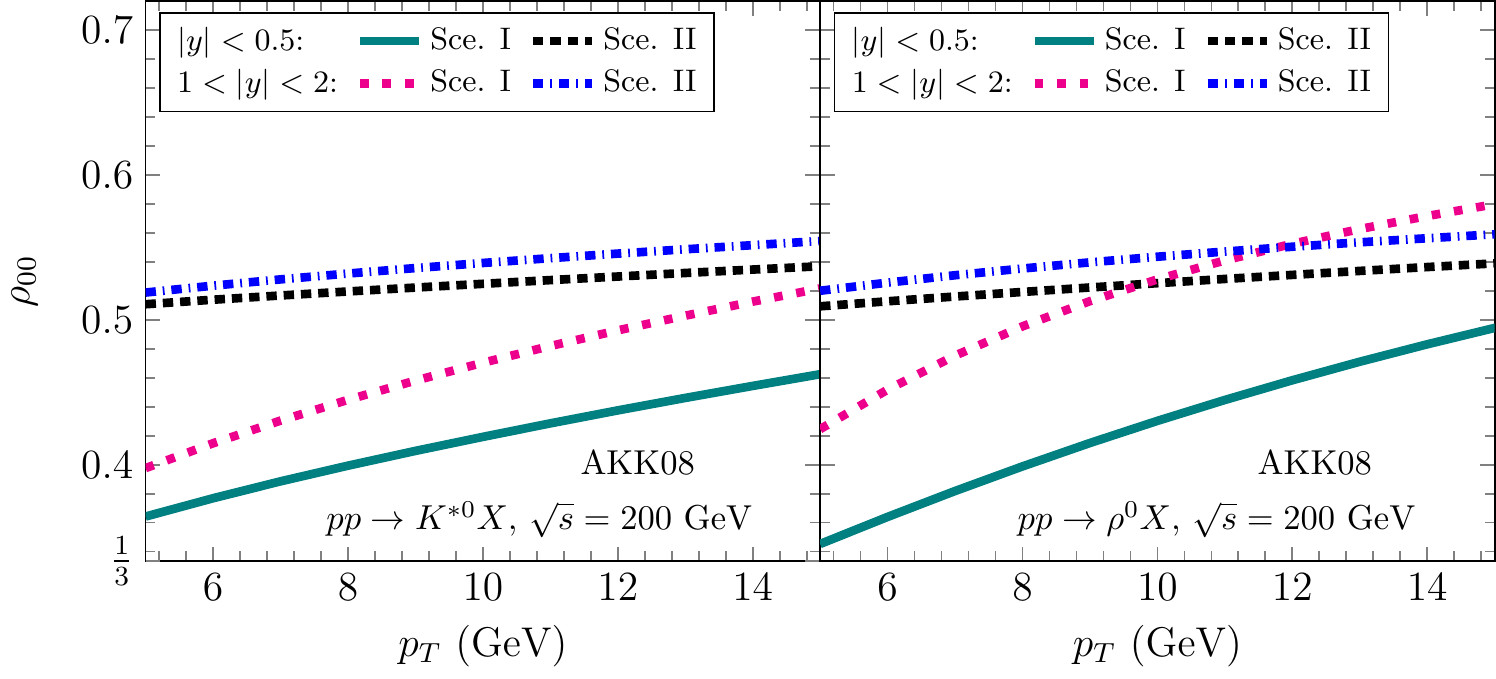}
\includegraphics[width=0.475\textwidth]{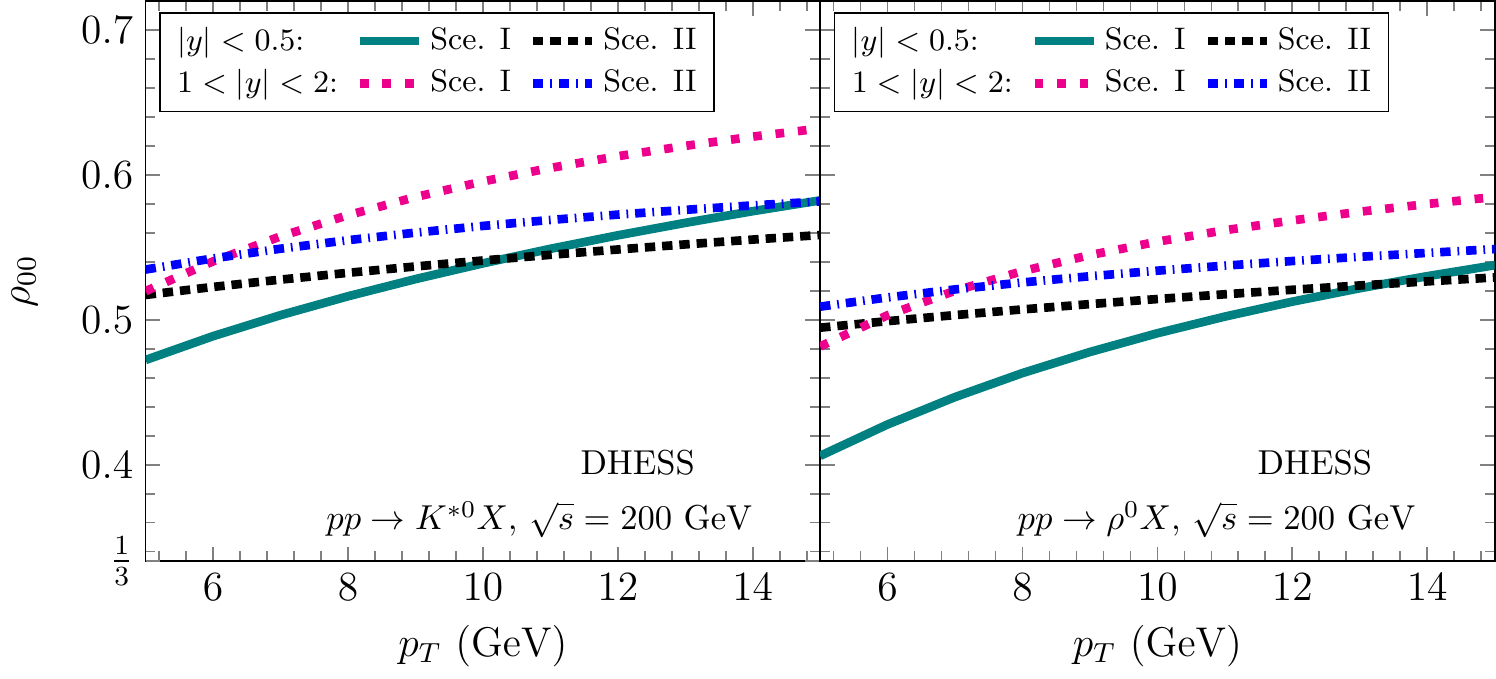}
\caption{(Color online) Spin alignments of vector mesons in $pp$ collisions at RHIC energy $\sqrt{s}=200$ GeV 
for $K^{*0}$ and $\rho^0$ in two rapidity regions as functions of $p_T$.}
\label{fig:SAatRHIC1}
\end{figure}

In Fig.~\ref{fig:SAatRHIC1}, we show the spin alignments for $K^{*0}$ and $\rho^0$  
at the RHIC energy in two rapidity regions as functions of $p_T$. 
From Fig.~\ref{fig:SAatRHIC1}, we see the following distinct features for the spin alignment in $pp\to VX$ at RHIC energy.

First, both the results for $K^{*0}$ and those for $\rho^0$ are significantly different from $1/3$, i.e., 
they show quite significant spin alignments in both cases. 
The deviations of $\rho_{00}$ from $1/3$ increase monotonically with increasing $p_T$. 
This is just consistent with the qualitative expectation mentioned at the end of Sec.~\ref{sec:Numpp1}. 
The increases with increasing $p_T$ are mainly due to increasing relative contributions from the quark fragmentation 
in particular those in the large $z$ region where $D_{1LL}/D_{1}$ is more significant.
We emphasize that these qualitative features are essential properties of the results and they are independent of the details of 
the parameterizations of the polarized FFs and unpolarized FFs.

Second, there is a significant difference between the results of the precise magnitudes obtained in scenario I and those in scenario II 
and there is also a quite obvious difference between results obtained using the two different sets of unpolarized FFs.
We see in particular a large difference between the results obtained with AKK08 and DHESS FFs in scenario I.  
These differences are mainly due to the large difference in gluon fragmentation functions in the two scenarios 
and in the two sets of unpolarized FFs.  
With AKK08, the gluon contribution to vector meson production is very large (see Fig.~\ref{fig:rVp2}) and this leads to 
a large difference between the results obtained with polarized FFs in scenario I and those in scenario II (upper panel in Fig.~\ref{fig:SAatRHIC1});   
while with DHESS, the gluon contribution is much smaller than that with AKK08 (see Fig.~\ref{fig:rVp2}), 
the difference becomes much smaller (lower panel in Fig.~\ref{fig:SAatRHIC1}). 

Third, there is also a quite significant difference between the results obtained in the two different rapidity regions. 
This is mainly because there are more contributions from the quark jets in the forward/backward rapidity. This leads to a larger vector meson spin alignment in a more forward/backward rapidity.

Forth, there is no distinct difference between the results for $K^{*0}$ and those for $\rho^0$. 
This is because that we have not considered the flavor dependence in our parameterizations of $D_{1LL}/D_1$. 
The small difference comes mainly from strangeness suppression in the unpolarized fragmentation functions. 

\begin{figure}[h!]
\includegraphics[width=0.475\textwidth]{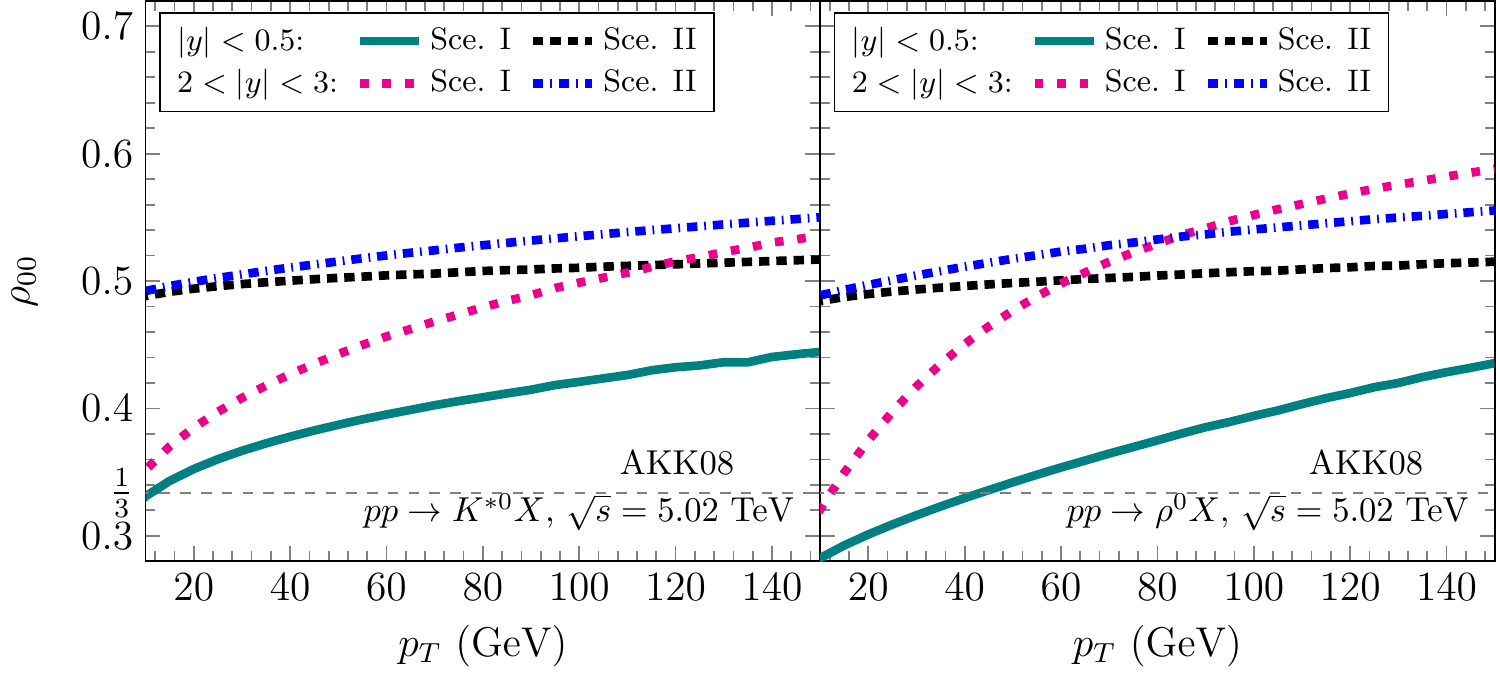}
\includegraphics[width=0.475\textwidth]{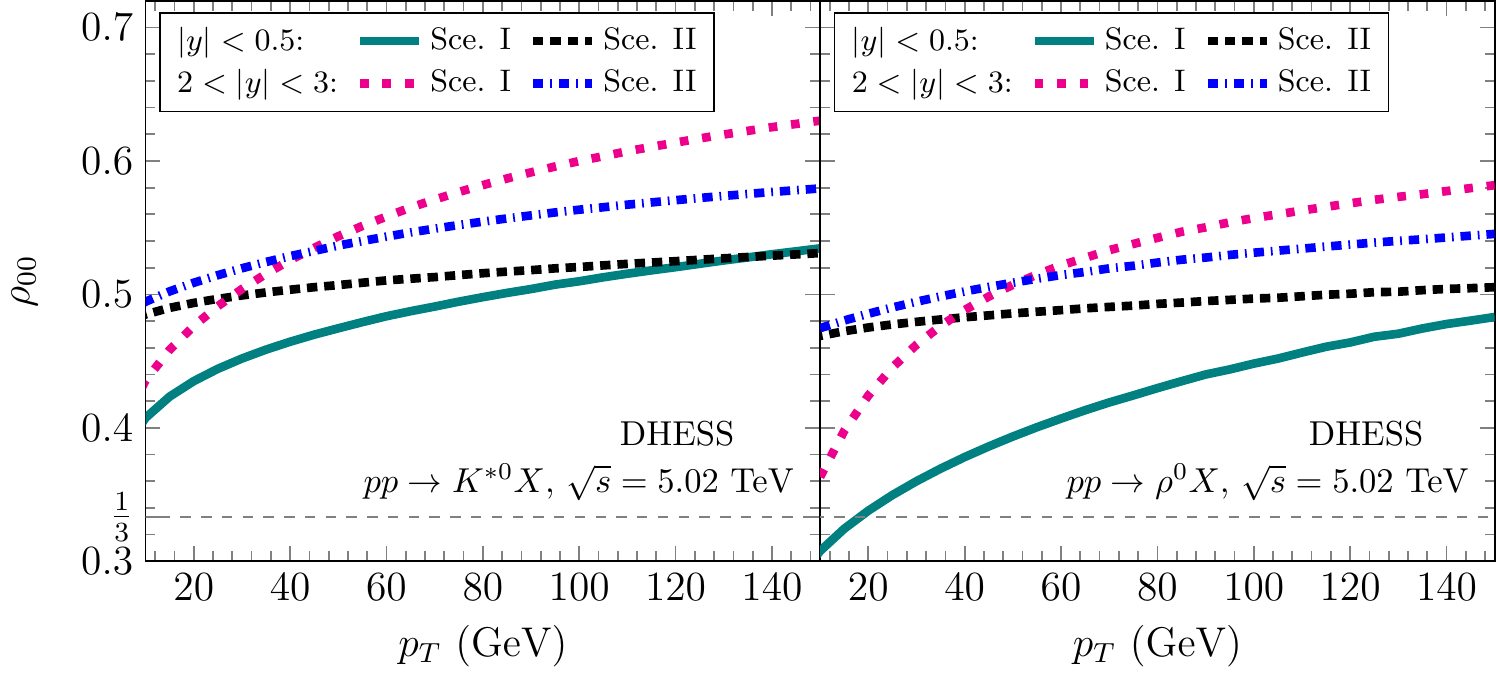}
\caption{(Color online) Spin alignments of vector mesons in $pp$ collisions at the LHC energy $\sqrt{s}=5.02$ TeV 
for $K^{*0}$ and $\rho^0$ in two rapidity regions as functions of $p_T$.}
\label{fig:SAatLHC1}
\end{figure}

In Fig.~\ref{fig:SAatLHC1}, we show the results obtained at the LHC energy. 
From Fig.~\ref{fig:SAatLHC1}, we see quite similar qualitative features as those seen from Fig.~\ref{fig:SAatRHIC1} at the RHIC energy. 
Here, we have the advantage to study a much wider $p_T$ range so that we can study the $p_T$-dependence more intensively. 
As mentioned above, the increase with $p_T$ of the spin alignment is caused by the increasing contributions from 
quarks fragmentations relative to the gluon fragmentation. 
It is also because the gluon contribution becomes more dominate at LHC energy in the relative small $p_T$ region in Fig.~\ref{fig:SAatLHC1},  
the spin alignment in that region is closer to $1/3$ and the differences between scenario I and II are also more significant. 

\begin{figure}[h!]
\includegraphics[width=0.475\textwidth]{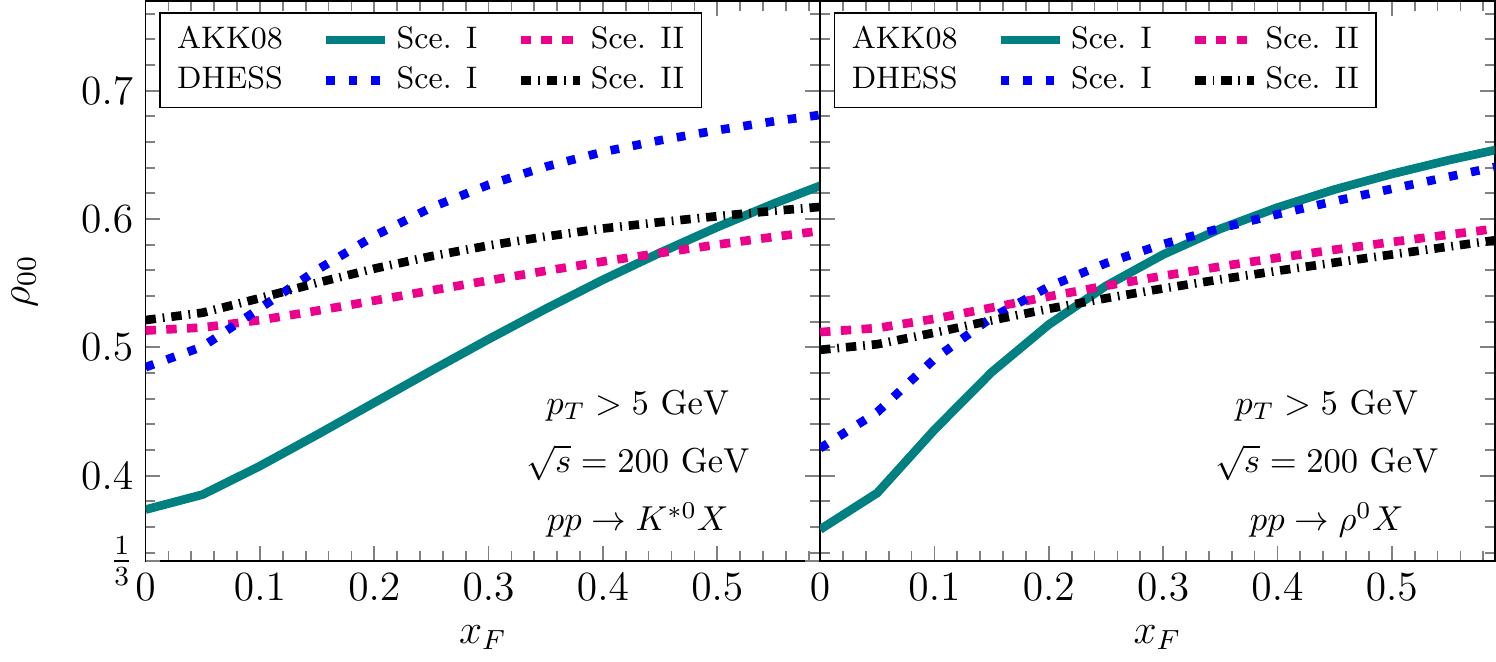}
\caption{(Color online) Spin alignments of vector mesons in $pp$ collisions at the RHIC energy $\sqrt{s}=200$ GeV 
for $K^{*0}$ and $\rho^0$ at $p_T>5$ GeV as functions of $x_F$.}
\label{fig:SAatRHIC2}
\end{figure}

\begin{figure}[h!]
\includegraphics[width=0.475\textwidth]{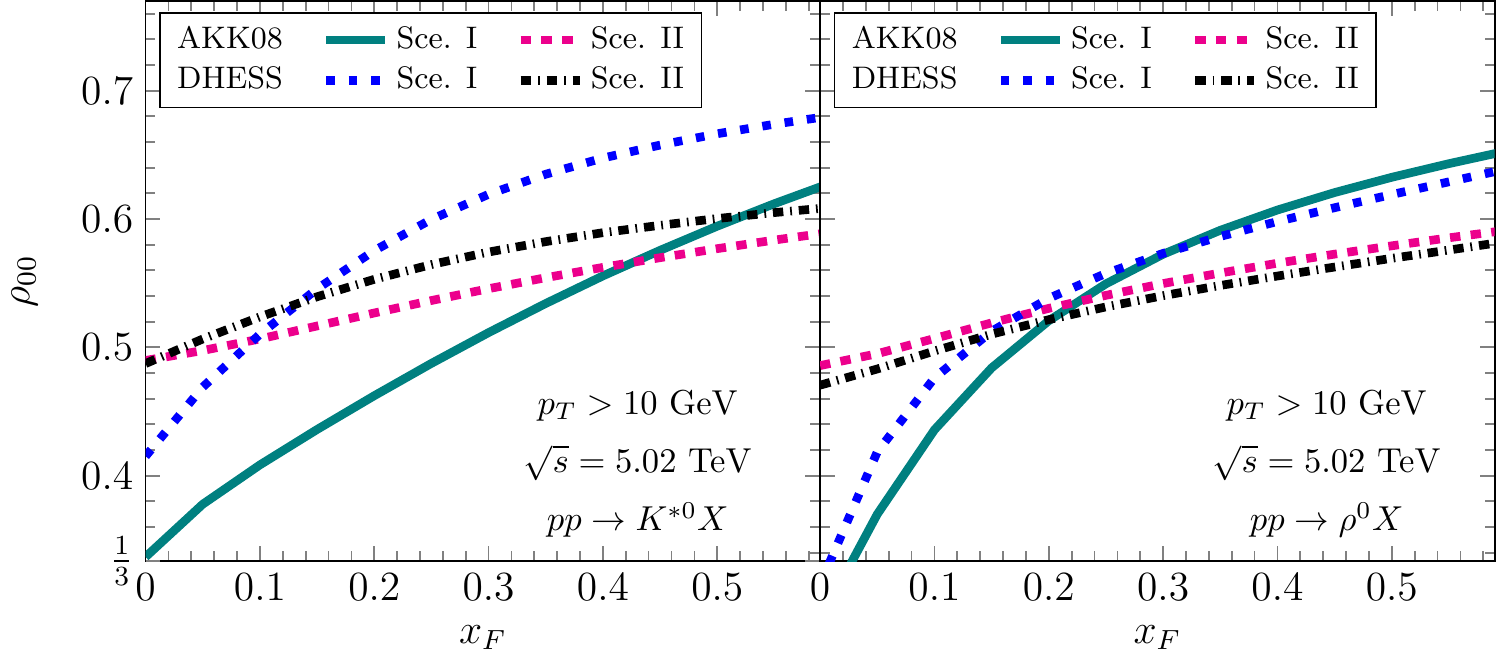}
\caption{(Color online) Spin alignments of vector mesons in $pp$ collisions at the LHC energy $\sqrt{s}=5.02$ TeV 
for $K^{*0}$ and $\rho^0$ at $p_T>10$ GeV as functions of $x_F$.}
\label{fig:SAatLHC2}
\end{figure}

In Figs.~\ref{fig:SAatRHIC2} and \ref{fig:SAatLHC2}, we show results for 
 $p_T$-integrated spin alignments of $K^{*0}$ and $\rho^0$
as functions of $x_F$ at RHIC and LHC energies respectively.

Here, from Figs.~\ref{fig:SAatRHIC2} and \ref{fig:SAatLHC2}, we see rapid increases of the spin alignment with increasing $x_F$, 
quite similar to that observed in $e^+e^-$ shown in Fig.~\ref{fig:fit-k-s1} 
and such a behavior is more obvious in scenario I. 
The increase reflects again the increasing relative contributions from favored quark fragmentations to the gluon fragmentation 
and also $z$-dependence of the favored $S_{LL}$-dependent FF $D_{1LL}$ relative to the corresponding unpolarized FF $D_1$. 
The relative larger values in the small $x_F$ region in scenario II are due to the quite large $D_{1LL}$ of gluon fragmentation in the small $z$ region. 
We recall that gluon fragmentation is even less known in the unpolarized case,  this provides also a good opportunity 
to study gluon fragmentation mechanism.
 
From all the results shown in Figs.~\ref{fig:SAatRHIC1}-\ref{fig:SAatLHC2}, we see clearly that 
spin alignments of vector mesons are in general quite significant in $pp\to VX$ at high energies. 
Studying these spin alignments should provide a good test to QCD fragmentation mechanism in general 
and differentiate between different parameterizations scenarios, 
provide precise information on quark or gluon fragmentation in different kinematic regions in particular.

\section{Summary}

In the QCD description of high energy reactions, the spin alignment of vector meson in a fragmentation process 
is described by the $S_{LL}$-dependent fragmentation function $D_{1LL}$ 
defined via the Lorentz decomposition of the quark-quark correlator. 
A systematic study of the Lorentz decomposition show that $D_{1LL}$ is independent of the polarization of the fragmenting quark. 
The first attempt to extract $D_{1LL}$ for $K^{*0}$ from the LEP data~\cite{Abreu:1997wd,Ackerstaff:1997kj} 
on $e^+e^-$-annihilations has been made in~\cite{Chen:2016iey}.

In this paper, we follow the same procedure of~\cite{Chen:2016iey} and make parameterizations of $D_{1LL}$ in two different scenarios  
for $K^{*0}$ and $\rho^0$ from different flavors of quarks, anti-quarks and gluon and evolve them using DGLAP equation. 
We apply the results obtained to $pp\to VX$ and make predictions for the spin alignment of vector mesons at RHIC and LHC energies.

The results show that the data~\cite{Abreu:1997wd,Ackerstaff:1997kj} available is far from enough to determine 
the precise forms of $D_{1LL}$ for different vector mesons from different flavors of quarks, anti-quarks and gluon. 
Nevertheless, we predict very significant spin alignments for vector mesons in $pp$ collisions at high energies. 
The results show a number of distinct features 
so that measurements of vector meson spin alignments in different kinematic regions in $pp$ collisions are 
not only able to check the quark polarization independence of $D_{1LL}$ 
but also sensitive to study the favored quark fragmentation and/or gluon fragmentation respectively.

\section*{Acknowledgements}
This work was supported in part by the National Natural Science Foundation of China
(approval Nos. 11675092, 11947055, 11890713, 11505080) and Shandong Province Natural Science Foundation Grant No. ZR2018JL006.


\begin{thebibliography}{99}




\bibitem{Abreu:1997wd} 
  P.~Abreu {\it et al.} [DELPHI Collaboration],
  Phys.\ Lett.\ B {\bf 406}, 271 (1997).
  doi:10.1016/S0370-2693(97)00758-2

\bibitem{Ackerstaff:1997kj} 
  K.~Ackerstaff {\it et al.} [OPAL Collaboration],
  Phys.\ Lett.\ B {\bf 412}, 210 (1997)
  doi:10.1016/S0370-2693(97)01077-0
  [hep-ex/9708022].

\bibitem{Abbiendi:1999bz} 
  G.~Abbiendi {\it et al.} [OPAL Collaboration],
  Eur.\ Phys.\ J.\ C {\bf 16}, 61 (2000)
  doi:10.1007/s100520050003
  [hep-ex/9906043].

\bibitem{Ackerstaff:1997kd} 
  K.~Ackerstaff {\it et al.}  [OPAL Collaboration],
  Z.\ Phys.\ C {\bf 74}, 437 (1997).

\bibitem{Chukanov:2006xa} 
  A.~Chukanov {\it et al.} [NOMAD Collaboration],
  Eur.\ Phys.\ J.\ C {\bf 46}, 69 (2006)
  doi:10.1140/EPJC/S2006-02500-4
  [hep-ex/0604050].


\bibitem{Abelev:2008ag} 
  B.~I.~Abelev {\it et al.} [STAR Collaboration],
  Phys.\ Rev.\ C {\bf 77}, 061902 (2008).


\bibitem{Zhou:2019lun} 
  C.~Zhou,
  Nucl.\ Phys.\ A {\bf 982}, 559 (2019).

\bibitem{Acharya:2019vpe} 
  S.~Acharya {\it et al.} [ALICE Collaboration],
  arXiv:1910.14408 [nucl-ex].




\bibitem{Lesnik:1975my} 
  A.~Lesnik {\it et al.}, 
  Phys.\ Rev.\ Lett.\  {\bf 35}, 770 (1975).

\bibitem{Bunce:1976yb} 
  G.~Bunce {\it et al.}, 
  Phys.\ Rev.\ Lett.\  {\bf 36}, 1113 (1976).

\bibitem{Bensinger:1983vc} 
  J.~Bensinger {\it et al.}, 
  Phys.\ Rev.\ Lett.\  {\bf 50}, 313 (1983).

\bibitem{Gourlay:1986mf} 
  S.~A.~Gourlay {\it et al.}, 
  Phys.\ Rev.\ Lett.\  {\bf 56}, 2244 (1986).

\bibitem{Adamovich:1994gy} 
  M.~I.~Adamovich {\it et al.} [WA89 Collaboration],
  Z.\ Phys.\ A {\bf 350}, 379 (1995)
  doi:10.1007/BF01291194
  [hep-ex/9409001].



\bibitem{Althoff:1984iz} 
  M.~Althoff {\it et al.} [TASSO Collaboration],
  Z.\ Phys.\ C {\bf 27}, 27 (1985).
  doi:10.1007/BF01642477


\bibitem{Buskulic:1996vb} 
  D.~Buskulic {\it et al.} [ALEPH Collaboration],
  Phys.\ Lett.\ B {\bf 374}, 319 (1996).
  doi:10.1016/0370-2693(96)00300-0

\bibitem{Ackerstaff:1997nh} 
  K.~Ackerstaff {\it et al.} [OPAL Collaboration],
  Eur.\ Phys.\ J.\ C {\bf 2}, 49 (1998)
  doi:10.1007/s100520050123
  [hep-ex/9708027].


\bibitem{Airapetian:1999sh} 
  A.~Airapetian {\it et al.} [HERMES Collaboration],
  Phys.\ Rev.\ D {\bf 64}, 112005 (2001)
  doi:10.1103/PhysRevD.64.112005
  [hep-ex/9911017].
  
\bibitem{Airapetian:2006ee} 
  A.~Airapetian {\it et al.} [HERMES Collaboration],
  Phys.\ Rev.\ D {\bf 74}, 072004 (2006) 
  doi:10.1103/PhysRevD.74.072004
  [hep-ex/0607004].

\bibitem{Astier:2000ax} 
  P.~Astier {\it et al.} [NOMAD Collaboration],
  Nucl.\ Phys.\ B {\bf 588}, 3 (2000).
  doi:10.1016/S0550-3213(00)00503-4

\bibitem{Astier:2001ve} 
  P.~Astier {\it et al.} [NOMAD Collaboration],
  Nucl.\ Phys.\ B {\bf 605}, 3 (2001)
  doi:10.1016/S0550-3213(01)00181-X
  [hep-ex/0103047].
  
\bibitem{Alekseev:2009ab} 
  M.~Alekseev {\it et al.} [COMPASS Collaboration],
  Eur.\ Phys.\ J.\ C {\bf 64}, 171 (2009)
  doi:10.1140/epjc/s10052-009-1143-7
  [arXiv:0907.0388 [hep-ex]].


\bibitem{Abelev:2009xg} 
  B.~I.~Abelev {\it et al.} [STAR Collaboration],
  Phys.\ Rev.\ D {\bf 80}, 111102 (2009)
  doi:10.1103/PhysRevD.80.111102
  [arXiv:0910.1428 [hep-ex]].

\bibitem{Adam:2018kzl} 
  J.~Adam {\it et al.} [STAR Collaboration],
  Phys.\ Rev.\ D {\bf 98}, 112009 (2018)
  doi:10.1103/PhysRevD.98.112009
  [arXiv:1808.07634 [hep-ex]].

\bibitem{Adam:2018wce} 
  J.~Adam {\it et al.} [STAR Collaboration],
  Phys.\ Rev.\ D {\bf 98}, 091103 (2018)
  doi:10.1103/PhysRevD.98.091103
  [arXiv:1808.08000 [hep-ex]].

\bibitem{Guan:2018ckx} 
  Y.~Guan {\it et al.} [Belle Collaboration],
  Phys.\ Rev.\ Lett.\  {\bf 122}, 042001 (2019)
  doi:10.1103/PhysRevLett.122.042001
  [arXiv:1808.05000 [hep-ex]].

\bibitem{BESIII}
M.~Ablikim {\it et al.} [BES III Collaboration], 
Nat. Phys. 15, 631–634 (2019). https://doi.org/10.1038/s41567-019-0494-8.


\bibitem{Gustafson:1992iq} 
  G.~Gustafson and J.~Hakkinen,
  Phys.\ Lett.\ B {\bf 303}, 350 (1993).
  doi:10.1016/0370-2693(93)91444-R

\bibitem{Liang:1997rt} 
  Z.~t.~Liang and C.~Boros,
  Phys.\ Rev.\ Lett.\  {\bf 79}, 3608 (1997)
  doi:10.1103/PhysRevLett.79.3608
  [hep-ph/9708488].

\bibitem{Boros:1998kc} 
  C.~Boros and Z.~t.~Liang,
  Phys.\ Rev.\ D {\bf 57}, 4491 (1998)
  doi:10.1103/PhysRevD.57.4491
  [hep-ph/9803225].

\bibitem{Liu:2000fi} 
  C.~x.~Liu and Z.~t.~Liang,
  Phys.\ Rev.\ D {\bf 62}, 094001 (2000)
  doi:10.1103/PhysRevD.62.094001
  [hep-ph/0005172].

\bibitem{Liu:2001yt} 
  C.~x.~Liu, Q.~h.~Xu and Z.~t.~Liang,
  Phys.\ Rev.\ D {\bf 64}, 073004 (2001)
  doi:10.1103/PhysRevD.64.073004
  [hep-ph/0106184].

\bibitem{Zuotang:2002ub} 
  Liang~Zuo-tang and Liu~Chun-xiu,
  Phys.\ Rev.\ D {\bf 66}, 057302 (2002).
  doi:10.1103/PhysRevD.66.057302

\bibitem{Xu:2002hz} 
  Q.~h.~Xu, C.~x.~Liu and Z.~t.~Liang,
  Phys.\ Rev.\ D {\bf 65}, 114008 (2002)
  doi:10.1103/PhysRevD.65.114008
  [hep-ph/0204318].

\bibitem{Dong:2005ea} 
  H.~Dong, J.~Zhou and Z.~t.~Liang,
  Phys.\ Rev.\ D {\bf 72}, 033006 (2005)
  doi:10.1103/PhysRevD.72.033006
  [hep-ph/0506207].

\bibitem{Xu:2005ru} 
  Q.~h.~Xu, Z.~t.~Liang and E.~Sichtermann,
  Phys.\ Rev.\ D {\bf 73}, 077503 (2006)
  doi:10.1103/PhysRevD.73.077503
  [hep-ph/0511061].

\bibitem{Chen:2007tm} 
  Y.~Chen, Z.~t.~Liang, E.~Sichtermann, Q.~h.~Xu and S.~s.~Zhou,
  Phys.\ Rev.\ D {\bf 78}, 054007 (2008)
  doi:10.1103/PhysRevD.78.054007
  [arXiv:0707.0534 [hep-ph]].

\bibitem{Zhou:2008fb} 
  J.~Zhou, F.~Yuan and Z.~T.~Liang,
  Phys.\ Rev.\ D {\bf 78}, 114008 (2008)
  doi:10.1103/PhysRevD.78.114008
  [arXiv:0808.3629 [hep-ph]].

\bibitem{Zhou:2009mx} 
  S.~s.~Zhou, Y.~Chen, Z.~t.~Liang and Q.~h.~Xu,
  Phys.\ Rev.\ D {\bf 79}, 094018 (2009)
  doi:10.1103/PhysRevD.79.094018
  [arXiv:0902.1883 [hep-ph]].


\bibitem{Ma:1998pd} 
  B.~Q.~Ma and J.~Soffer,
  Phys.\ Rev.\ Lett.\  {\bf 82}, 2250 (1999)
  doi:10.1103/PhysRevLett.82.2250
  [hep-ph/9810517].


\bibitem{Ma:1999gj} 
  B.~Q.~Ma, I.~Schmidt and J.~J.~Yang,
  Phys.\ Lett.\ B {\bf 477}, 107 (2000)
  doi:10.1016/S0370-2693(00)00167-2
  [hep-ph/9906424].


\bibitem{Ma:1999wp} 
  B.~Q.~Ma, I.~Schmidt and J.~J.~Yang,
  Phys.\ Rev.\ D {\bf 61}, 034017 (2000)
  doi:10.1103/PhysRevD.61.034017
  [hep-ph/9907224].


\bibitem{Ma:1999hi} 
  B.~Q.~Ma, I.~Schmidt and J.~J.~Yang,
  Nucl.\ Phys.\ B {\bf 574}, 331 (2000)
  doi:10.1016/S0550-3213(00)00021-3
  [hep-ph/9907556].


\bibitem{Ma:2000uu} 
  B.~Q.~Ma, I.~Schmidt, J.~Soffer and J.~J.~Yang,
  Eur.\ Phys.\ J.\ C {\bf 16}, 657 (2000)
  doi:10.1007/s100520000447
  [hep-ph/0001259].


\bibitem{Ma:2000cg} 
  B.~Q.~Ma, I.~Schmidt, J.~Soffer and J.~J.~Yang,
  Phys.\ Rev.\ D {\bf 62}, 114009 (2000)
  doi:10.1103/PhysRevD.62.114009
  [hep-ph/0008295].


\bibitem{Chi:2013hka} 
  Y.~Chi and B.~Q.~Ma,
  Phys.\ Lett.\ B {\bf 726}, 737 (2013)
  doi:10.1016/j.physletb.2013.09.044
  [arXiv:1310.2005 [hep-ph]].

\bibitem{Liu:2019xcf} 
  X.~Liu and B.~Q.~Ma,
  Eur.\ Phys.\ J.\ C {\bf 79}, no. 5, 409 (2019)
  doi:10.1140/epjc/s10052-019-6916-z
  [arXiv:1905.02360 [hep-ph]].


\bibitem{Ellis:2002zv} 
  J.~R.~Ellis, A.~Kotzinian and D.~V.~Naumov,
  Eur.\ Phys.\ J.\ C {\bf 25}, 603 (2002)
  doi:10.1140/epjc/s2002-01025-2
  [hep-ph/0204206].


\bibitem{deFlorian:1997zj} 
  D.~de Florian, M.~Stratmann and W.~Vogelsang,
 Phys.\ Rev.\ D {\bf 57}, 5811 (1998)
 doi:10.1103/PhysRevD.57.5811
  [hep-ph/9711387].


\bibitem{Anselmino:1984af} 
  M.~Anselmino and P.~Kroll,
  Phys.\ Rev.\ D {\bf 30}, 36 (1984).



\bibitem{Anselmino:1997ui} 
  M.~Anselmino, M.~Bertini, F.~Murgia and P.~Quintairos,
  Eur.\ Phys.\ J.\ C {\bf 2}, 539 (1998)
  doi:10.1007/s100520050159
  [hep-ph/9704420].


\bibitem{Anselmino:1998jv} 
  M.~Anselmino, M.~Bertini, F.~Murgia and B.~Pire,
  Phys.\ Lett.\ B {\bf 438}, 347 (1998)
  doi:10.1016/S0370-2693(98)00978-2
  [hep-ph/9805234].


\bibitem{Anselmino:1999cg} 
  M.~Anselmino, M.~Bertini, F.~Caruso, F.~Murgia and P.~Quintairos,
  Eur.\ Phys.\ J.\ C {\bf 11}, 529 (1999)
  doi:10.1007/s100529900200, 10.1007/s100520050652
  [hep-ph/9904205].


\bibitem{Xu:2001hz} 
  Q.~h.~Xu, C.~x.~Liu and Z.~t.~Liang,
  Phys.\ Rev.\ D {\bf 63}, 111301 (2001)
  doi:10.1103/PhysRevD.63.111301
  [hep-ph/0103267].


\bibitem{Xu:2002vz} 
  Q.~h.~Xu and Z.~t.~Liang,
  Phys.\ Rev.\ D {\bf 66}, 017301 (2002)
  doi:10.1103/PhysRevD.66.017301
  [hep-ph/0205291].

\bibitem{Xu:2003fq} 
  Q.~h.~Xu and Z.~t.~Liang,
  Phys.\ Rev.\ D {\bf 67}, 114013 (2003)
  doi:10.1103/PhysRevD.67.114013
  [hep-ph/0304125].

\bibitem{Xu:2003rs} 
  Q.~h.~Xu and Z.~t.~Liang,
  Phys.\ Rev.\ D {\bf 68}, 034023 (2003)
  doi:10.1103/PhysRevD.68.034023
  [hep-ph/0307327].

  
\bibitem{Boer:1997mf} 
  D.~Boer, R.~Jakob and P.~J.~Mulders,
  Nucl.\ Phys.\ B {\bf 504}, 345 (1997)
  doi:10.1016/S0550-3213(97)00456-2
  [hep-ph/9702281].

\bibitem{Boer:1997qn} 
  D.~Boer, R.~Jakob and P.~J.~Mulders,
  Phys.\ Lett.\ B {\bf 424}, 143 (1998)
  doi:10.1016/S0370-2693(98)00136-1
  [hep-ph/9711488].

\bibitem{Boer:2008fr} 
  D.~Boer,
  Nucl.\ Phys.\ B {\bf 806}, 23 (2009)
  doi:10.1016/ j.nuclphysb.2008.06.011
  [arXiv:0804.2408[hep-ph]].

\bibitem{Pitonyak:2013dsu} 
  D.~Pitonyak, M.~Schlegel and A.~Metz,
  Phys.\ Rev.\ D {\bf 89}, no. 5, 054032 (2014)
  doi:10.1103/PhysRevD.89.054032
  [arXiv:1310.6240 [hep-ph]].


\bibitem{Wei:2013csa} 
  S.~y.~Wei, Y.~k.~Song and Z.~t.~Liang,
  Phys.\ Rev.\ D {\bf 89}, no. 1, 014024 (2014)
  doi:10.1103/PhysRevD.89.014024
  [arXiv:1309.4191 [hep-ph]].

\bibitem{Wei:2014pma} 
  S.~Y.~Wei, K.~b.~Chen, Y.~k.~Song and Z.~t.~Liang,
  Phys.\ Rev.\ D {\bf 91}, no. 3, 034015 (2015)
  doi:10.1103/PhysRevD.91.034015
  [arXiv:1410.4314 [hep-ph]].

\bibitem{Chen:2015ora} 
  K.~b.~Chen, S.~y.~Wei, W.~h.~Yang and Z.~t.~Liang,
  arXiv:1505.02856 [hep-ph].

\bibitem{Chen:2016moq} 
  K.~b.~Chen, W.~h.~Yang, S.~y.~Wei and Z.~t.~Liang,
  Phys.\ Rev.\ D {\bf 94}, no. 3, 034003 (2016)
  doi:10.1103/PhysRevD.94.034003
  [arXiv:1605.07790 [hep-ph]].

\bibitem{Chen:2016iey} 
  K.~b.~Chen, W.~h.~Yang, Y.~j.~Zhou and Z.~t.~Liang,
  Phys.\ Rev.\ D {\bf 95}, no. 3, 034009 (2017)
  doi:10.1103/PhysRevD.95.034009
  [arXiv:1609.07001 [hep-ph]].

\bibitem{Wei:2016far} 
  S.~y.~Wei, Y.~k.~Song, K.~b.~Chen and Z.~t.~Liang,
  Phys.\ Rev.\ D {\bf 95}, no. 7, 074017 (2017)
  doi:10.1103/PhysRevD.95.074017
  [arXiv:1611.08688 [hep-ph]].

\bibitem{Chen:2015tca} 
  K.~b.~Chen, S.~y.~Wei and Z.~t.~Liang,
  Front.\ Phys.\ (Beijing) {\bf 10}, no. 6, 101204 (2015)
  doi:10.1007/s11467-015-0477-x
  [arXiv:1506.07302 [hep-ph]].

 
\bibitem{Liang:2004ph} 
  Z.~T.~Liang and X.~N.~Wang,
  Phys.\ Rev.\ Lett.\  {\bf 94}, 102301 (2005)
  Erratum: [Phys.\ Rev.\ Lett.\  {\bf 96}, 039901 (2006)]. 
  doi:10.1103/PhysRevLett.94.102301, 10.1103/PhysRevLett.96.039901
  [nucl-th/0410079].

\bibitem{Liang:2004xn} 
  Z.~T.~Liang and X.~N.~Wang,
  Phys.\ Lett.\ B {\bf 629}, 20 (2005)
  doi:10.1016/j.physletb.2005.09.060
  [nucl-th/0411101].

\bibitem{STAR:2017ckg} 
  L.~Adamczyk {\it et al.} [STAR Collaboration],
  Nature {\bf 548}, 62 (2017). 
  doi:10.1038/nature23004. 
  [arXiv:1701.06657 [nucl-ex]].
  
  
 \bibitem{Bacchetta:2000jk} 
  A.~Bacchetta and P.~J.~Mulders,
  Phys.\ Rev.\ D {\bf 62}, 114004 (2000)
  [hep-ph/0007120].


\bibitem{Owens:1986mp} 
  J.~F.~Owens,
  Rev.\ Mod.\ Phys.\  {\bf 59}, 465 (1987).
  doi:10.1103/RevModPhys.59.465

\bibitem{Dulat:2015mca} 
  S.~Dulat {\it et al.},
  Phys.\ Rev.\ D {\bf 93}, no. 3, 033006 (2016)
  doi:10.1103/PhysRevD.93.033006
  [arXiv:1506.07443 [hep-ph]].

 
\bibitem{Dokshitzer:1977sg} 
  Y.~L.~Dokshitzer,
  Sov.\ Phys.\ JETP {\bf 46}, 641 (1977)
  [Zh.\ Eksp.\ Teor.\ Fiz.\  {\bf 73}, 1216 (1977)].

\bibitem{Gribov:1972ri} 
  V.~N.~Gribov and L.~N.~Lipatov,
  Sov.\ J.\ Nucl.\ Phys.\  {\bf 15}, 438 (1972)
  [Yad.\ Fiz.\  {\bf 15}, 781 (1972)].
  
\bibitem{Gribov:1972rt} 
  V.~N.~Gribov and L.~N.~Lipatov,
  Sov.\ J.\ Nucl.\ Phys.\  {\bf 15}, 675 (1972)
  [Yad.\ Fiz.\  {\bf 15}, 1218 (1972)].


\bibitem{Altarelli:1977zs} 
  G.~Altarelli and G.~Parisi,
  Nucl.\ Phys.\ B {\bf 126}, 298 (1977).
  doi:10.1016/0550-3213(77)90384-4

\bibitem{Owens:1978qz} 
  J.~F.~Owens,
  Phys.\ Lett.\ B {\bf 76}, 85 (1978).
  doi:10.1016/0370-2693(78)90108-9


\bibitem{Georgi:1977mg} 
  H.~Georgi and H.~D.~Politzer,
  Nucl.\ Phys.\ B {\bf 136}, 445 (1978).
  doi:10.1016/0550-3213(78)90269-9

\bibitem{Uematsu:1978yw} 
  T.~Uematsu,
  Phys.\ Lett.\ B {\bf 79}, 97 (1978).
  doi:10.1016/0370-2693(78)90444-6



\bibitem{Albino:2008fy} 
  S.~Albino, B.~A.~Kniehl and G.~Kramer,
  Nucl.\ Phys.\ B {\bf 803}, 42 (2008)
  doi:10.1016/j.nuclphysb.2008.05.017
  [arXiv:0803.2768 [hep-ph]].

\bibitem{deFlorian:2017lwf}
D.~de Florian, M.~Epele, R.~Hernandez-Pinto, R.~Sassot and M.~Stratmann,
Phys. Rev. D \textbf{95} (2017) no.9, 094019
doi:10.1103/PhysRevD.95.094019
[arXiv:1702.06353 [hep-ph]].


\bibitem{Shlyapnikov:2001jf} 
  P.~V.~Shlyapnikov,
  Phys.\ Lett.\ B {\bf 512}, 18 (2001)
  doi:10.1016/S0370-2693(01)00691-8
  [hep-ph/0112084].


 


\end{thebibliography}
\end{document}